\documentclass[twocolumn,superscriptaddress,showpacs,notitlepage]{revtex4-1}
\usepackage{amsmath,amssymb,amsfonts}
\usepackage{graphicx}
\usepackage{txfonts}
\usepackage{color}
\usepackage{hyperref}
\usepackage{framed}
\usepackage{wrapfig}
\usepackage[letterpaper]{geometry}
\usepackage{soul}
\usepackage[normalem]{ulem}
\definecolor{fgr}{rgb}{0.13, 0.55, 0.13}

 \definecolor{ashgrey}{rgb}{0.7, 0.75, 0.71}

\DeclareMathOperator\sgn{sgn}

\newcommand{\im}{\mathrm{i}}

\newcommand{\bk}{\boldsymbol k}
\newcommand{\br}{\boldsymbol r}

\newcommand{\e}{\varepsilon}

\begin{document}

\title{Dipole excitation of collective modes in viscous two-dimensional electron systems}

\author{Vera Andreeva}
\affiliation{School of Mathematics, University of Minnesota, Minneapolis 55455, USA}
\author{Denis A. Bandurin}
\affiliation{Department of Physics, Massachusetts Institute of Technology, Cambridge, Massachusetts 02139, USA}
\author{Mitchell Luskin}
\affiliation{School of Mathematics, University of Minnesota, Minneapolis 55455, USA}
\author{Dionisios Margetis}
\affiliation{Institute for Physical Science and Technology, and Department of Mathematics,
and Center for Scientific Computation and Mathematical Modeling, University of Maryland, College Park, Maryland 20742, USA}

\begin{abstract}
We describe the structure of the time-harmonic electromagnetic field of a vertical Hertzian electric dipole source radiating over an  infinite, translation invariant two-dimensional electron system. Our model for the electron flow takes into account the effects of shear and Hall viscosities as well as an external static magnetic field perpendicular to the sheet. We identify two wave modes, namely, a surface plasmon and a diffusive mode. In the presence of an  external static magnetic field, the diffusive mode combines the features of both the conventional and Hall diffusion and may exhibit a negative group velocity. In our analysis, we solve exactly a boundary value problem for the time-harmonic Maxwell equations coupled with linearized hydrodynamic equations for the flat, two-dimensional material.  By numerically evaluating the integrals for the electromagnetic field on the sheet, we find that the plasmon contribution dominates in the intermediate-field region of the dipole source. In contrast, the amplitude of the diffusive mode reaches its maximum value in the near-field region, and quickly decays with the distance from the source. We demonstrate that the diffusive mode can be distinguished from the plasmon in the presence of the static magnetic field, when the highly oscillatory plasmon is gapped and tends to disappear.
\end{abstract}
\maketitle

\section{Introduction}
\label{sec:Intro}
In metallic conductors with low disorder and weak electron-phonon coupling, the momentum-conserving electron-electron collisions render the local thermodynamic equilibrium ensuring a fluid-like electron behavior. In this situation, the linear size of the system is larger than the mean free path of
momentum-conserving scattering processes but much smaller than the mean free path for momentum-relaxing collisions.
The emerging electron behavior, predicted several decades ago~\cite{Gurzhi1968}, is readily described by a hydrodynamic theory that is similar to the one used to treat  transport phenomena in classical liquids and gases.
Recent experimental evidence for electronic hydrodynamics in solid-state materials such as GaAs~\cite{molenkamp1994observation, de1995hydrodynamic, BraemPRB2018}, PdCoO$_2$ ~\cite{moll2016evidence}, WP$_2$~\cite{gooth2017electrical} and graphene~\cite{bandurin2016negative, crossno2016observation, kumar2017superballistic, FluidOnset, berdyugin2019measuring, Gallagher158, ShahalPoi2020} has spiked active interest in the hydrodynamic approach within the linear and nonlinear responses. Special attention focuses on two spatial dimensions, where electron-electron interactions are particularly strong~\cite{giuliani2005quantum}.

As a macroscopic theory for strongly interacting particle systems, hydrodynamics is a valuable tool in the study of problems where strong correlations invalidate simple theoretical approaches based on single-particle considerations~\cite{muller2009graphene, AndKivSpivak2011, Torre2015, LF2016, AlekseevNegative, narozhny2019electronic}.
Interacting many-body quantum systems allow for a hydrodynamic description
when the typical electron-electron scattering time, $\tau_{ee}$, is the shortest time scale of the system~\cite{lucas2018hydrodynamics}. Therefore, when an external excitation in the form of an electromagnetic field at a sufficiently high frequency, $\omega$, is supplied to a two-dimensional electron system (2DES), the latter responds ``hydrodynamically'' if $\omega\ll\tau_{ee}^{-1}$. For example, in graphene this condition is satisfied in the THz and sub-THz frequency ranges at sufficiently elevated temperatures. Motivated by these developments, in this paper we systematically address the coupling of the hydrodynamic 2DES to the field generated by an elementary electric-current-carrying source.

It has been predicted that hydrodynamic 2DESs host a variety of intriguing collective modes. One can recall the conventional plasmon modes, which are characterized by a dispersion relation of the form $\omega\sim \sqrt{k}$ (where $k$ is the mode wavenumber) and morph into collisional plasma waves with $\omega\sim k$ if only short-range interactions are respected~\cite{Dima-crossover, lucas2016sound, lucas2018electronic, sun2018universal}. In addition to the conventional plasmon modes, hydrodynamic 2DESs are also expected to support the propagation of transverse shear waves~\cite{ShearGregory} and, in the presence of a static magnetic field, magnetosonic waves ~\cite{AlekseevMagnetosound, Counterflow}.
Furthermore, an exotic electron-hole sound has been predicted to emerge in the graphene electron fluid close to the charge neutrality point where particles of opposite types coexist~\cite{svintsov2012hydrodynamic, phan2013ballistic, SunDemons2016, sun2018universal}.
In consideration of the rich variety of collective modes in hydrodynamic 2DESs, it is natural to expect that the electromagnetic response of such systems is strongly affected by the excitations of these modes~\cite{EMpropZaanen, EM_hallvisc, sherafati2016hall, svintsov2013hydrodynamic}. If the incident field is produced by an electric-current-carrying source, these modes accompany a radiation field that has a complicated spatial structure.

In this paper, we describe the excitation of collective modes in a viscous 2DES by the high-frequency electromagnetic field generated by a Hertzian electric dipole. We demonstrate that the electromagnetic response of the electron fluid contains two fundamentally different collective modes, namely, a surface `plasmon' and a `diffusive' mode, which accompany the radiation field. The dispersion relation of the former mode (plasmon) primarily depends on the equilibrium electron density through a Drude-like weight. The amplitude of this mode is appreciable in the intermediate-field region of the dipole source even if the viscous effects tend to disappear. In contrast, the diffusive mode is characterized by a strong dependence on the viscosities of the electron fluid, and combines features of both the conventional and Hall diffusion. The amplitude of this mode is noticeable in the near-field region but quickly decays away from the source. We point out the possible scale separation in the manifestation of the two modes. Furthermore, by including a static magnetic field we describe an intriguing interplay of its direct effect with that of the Hall viscosity in the angular component of the electric field. 

In our model, the material sheet is infinite and translation invariant, and lies in an unbounded, homogeneous and isotropic dielectric medium. The dipole is vertical to the material boundary; and in principle excites all three electric field components in the presence of a static magnetic field perpendicular to the sheet. Because of the character of the source, the spatial behavior of the generated total field on the sheet ranges from that of a near field, which tends to be singular at the source, to an intermediate field and, for large enough distances from the dipole, to the distinctly different far field.
We study in detail this spatial structure in connection to the aforementioned collective modes. For this purpose, we solve a boundary value problem for the time-harmonic Maxwell equations coupled with linearized viscous hydrodynamic (Navier-Stokes-type) equations for the flow of the 2DES. In this analysis, we include a static magnetic field perpendicular to the plane of the 2DES.

Our choice of a linear hydrodynamic model can be justified via the following consideration. The observation of nonlinear electron hydrodynamics requires a conducting material with a large momentum relaxation time~\cite{lucas2018hydrodynamics}. This property is very difficult to achieve experimentally. Hence, we restrict our attention to linear hydrodynamics, using linearized Navier-Stokes-type equations in the 2D sheet. Although we incorporate viscous effects by some analogy with classical liquids and gases, we place emphasis on the special role of the Hall viscosity in the 2DES response.

We reiterate that our analysis and numerics demonstrate the coexistence of two distinct collective modes, the surface plasmon and diffusive mode, with the radiation field on the material sheet. To be more precise, these modes are analytically identified with the contributions of certain poles in the complex plane of the Fourier variable that corresponds to the (radial) wavevector component tangential to the sheet. In contrast, the radiation field is a contribution associated with (but not entirely determined by) a different type of singularity, and is characterized by the free-space wave number. The plasmon and the diffusive mode are affected by the \emph{nonlocal} character of the emerging surface conductivity of the sheet because of the underlying hydrodynamic behavior.
A highlight of our results is that the dispersion relations of these modes are distinctly different from those of the conventional (transverse-magnetic or transverse-electric polarized) graphene plasmons; in the latter modes, nonlocal effects in the surface conductivity are typically neglected. Notably, regarding the diffusive mode in the presence of an external magnetic field, we show that the Hall viscosity affects the dispersion relation of this mode significantly and can change the sign of its group velocity. Furthermore, we describe the range of distance from the dipole source for which each of the surface wave modes can be dominant.

To make a connection with previous surface conductivity models, a part of our analysis focuses on the derivation of the underlying nonlocal surface conductivity tensor as a function of the wavevector on the sheet and frequency. In view of the hydrodynamic ingredients of our model, this conductivity takes into account the nonlocal electrical transport characteristics of 2DESs. In our setting, these features include the shear and Hall viscosities~\cite{berdyugin2019measuring, pellegrino2017nonlocal} and fluid compressibility, and are subject to a perpendicular static magnetic field. By coupling the hydrodynamic description of the 2D material with time-harmonic Maxwell's equations,
we derive Sommerfeld-type integrals for the electric-field components. We numerically compute these integrals, and single out and assess the contributions of the plasmon and diffusive mode in comparison to the radiation field for a wide range of distance from the source.

Notably, we find that the electric-field amplitude of the diffusive mode in graphene \emph{peaks in the near-field region}. The diffusive mode can have a smaller wavelength than the plasmon, but its spatial decay due to dissipation is intrinsically stronger. Unlike the plasmon, the diffusive mode does not exhibit a gap in the dispersion relation when a static magnetic field is applied. This mode is, however, very sensitive to the nonlocal properties of the material. In contrast, the plasmon mode \emph{dominates in the intermediate-field region}. In addition, its dispersion relation can be gapped by a magnetic field, if the frequency is suitably chosen, and the nonlocal effect on the sheet primarily influences its dissipation. Our results provide predictions that may guide the detection of wave modes inherent to the hydrodynamic 2DES by use of electromagnetic probes such as scanning-type near field optical microscopy or antenna-coupled field-effect  transistors~\cite{bandurin2018resonant,Khavronin2020singularity}.

We should mention a few open questions motivated by our analysis. The use of more realistic electric-current-carrying sources such as a linear antenna of finite size is a tractable problem of experimental interest. In our hydrodynamic model, we invoke a simplified version of the linearized compressible Navier-Stokes equations with a Hall viscosity. This model can be enriched with more complicated constitutive laws~\cite{briskot2015collision,sun2018universal,sun2018thirdorder}. The character of the electromagnetic fields in the time domain, when the source radiates a pulse, is not touched upon here. This aspect should be developed if electromagnetism needs to be coupled with a broader range of hydrodynamic phenomena including the intrinsically nonlinear convective acceleration.

The remainder of the paper is organized as follows.
In Sec.~\ref{sec:model}, we introduce the linear hydrodynamic model coupled with Maxwell's equations for the problem at hand.
In Sec.~\ref{sec:conductivity}, we derive the nonlocal conductivity tensor for the material sheet which takes into account the effects of shear and Hall viscosities, compressibility and external static magnetic field.
In Sec.~\ref{sec:integrals}, we express the excited electromagnetic field by a Fourier-Bessel representation, also known as the ``Hankel transform''; and discuss qualitative features of this description.
Section~\ref{sec:poles} focuses on the derivation of the dispersion relations for the wave modes on the material sheet. This task involves the examination of singularities present in the Fourier-Bessel transforms of the electric-field components.
In Sec.\ref{sec:amplitudes}, we evaluate numerically the requisite integrals when both the dipole source and the observation
point lie in the material sheet. Section~\ref{sec:discussion} provides a brief discussion on implications of our results.
In Sec.~\ref{sec:conclusion} we conclude the paper with a summary of the main results.

We assume that the time-harmonic fields have the temporal dependence $e^{-\im\omega t}$, where $\omega$ is the angular frequency.
We use the centimetre-gram-second (CGS) system of units.

\section{Formulation}
\label{sec:model}

In this section, we describe the geometry and governing equations for the problem under consideration. In our formulation, we combine Maxwell's equations for the electromagnetic field with a linear viscous hydrodynamic model for the 2DES. The electronic fluid flow is modeled via linearized Navier-Stokes-type equations, which consist of the continuity and momentum equations with viscous effects.

\begin{figure}[h]
\includegraphics[width=0.7\linewidth]{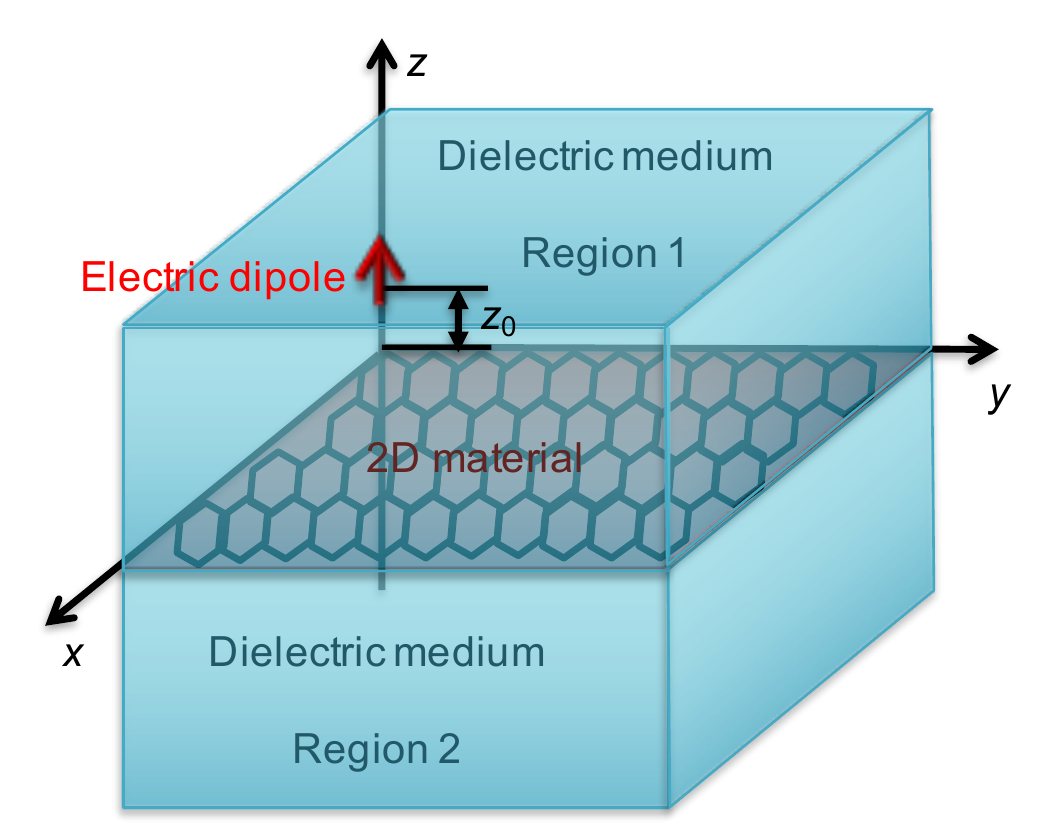}
\centering{}\caption{Geometry of the problem.
An infinite, flat conducting material sheet lies in the $xy$-plane, inside an unbounded dielectric medium. A vertical, $z$-directed, Hertzian electric dipole is located at distance $z_0$ from the sheet. The region with $z>0$ ($z<0$) is labeled by the index $j=1$ ($j=2$).}
\label{fig0}
\end{figure}

In our setting, the conducting sheet of the 2DES lies in the $xy$-plane, and is immersed in an unbounded linear, isotropic and homogeneous dielectric medium, as shown in Fig.~\ref{fig0}. A $z$-directed Hertzian electric dipole of unit moment is located at height $z_0$ above the sheet.
The ambient dielectric medium has permittivity relative to the vacuum equal to $\e$ (for $z\neq 0$). We label the region of the upper (lower) half space, for $z>0$ ($z<0$), by the index $j=1$ ($j=2$).

We start with the time-harmonic Maxwell equations along with suitable (transmission) boundary conditions for the electromagnetic field on the material sheet. These boundary conditions account for a surface current density at $z=0$ due to the charge flow in the 2DES~\cite{Bludov2013}.

The curl laws of Maxwell's equations in region $j$ are
\begin{subequations}\label{eqs:curl-M}
\begin{align}
\label{eq:m1}
\nabla\times\bold{H}_{j}&=-\frac{\im\omega}{c}\bold{D}_{j}+\frac{4\pi}{c}\bold{J},\\
\label{eq:m2}
\nabla\times\bold{E}_{j}&=\frac{\im\omega}{c}\bold{H}_{j}\qquad (j=1,\,2).
\end{align}
\end{subequations}
In the above, $\bold{E}_{j}\left(x,y,z\right)$, $\bold{H}_{j}\left(x,y,z\right)$ and $\bold{D}_{j}\left(x,y,z\right)=\e\bold{E}_{j}\left(x,y,z\right)$ are the electric, magnetic and displacement vector fields, respectively. We define
\begin{align*}
\bold{J}\left(x,y,z\right)=(I_0\ell)\delta\left(x\right)\delta\left(y\right)\delta\left(z-z_0\right)\, \bold{e}_z
\end{align*}
as the vector-valued volume current density due to the vertical electric dipole. Here, $I_0\ell$ denotes the dipole strength (electric moment) where $I_0$ and $\ell$ have units of current and length, respectively; $z_0$ is the height of the dipole above the material sheet ($z_0>0$ for definiteness); $c$ is the speed of light in vacuum; $\bold{e}_z$ is the $z$-directed unit Cartesian vector; and $\delta(x)$ is Dirac's delta function in one dimension. For a dipole of unit electric moment we
    set $I_0\ell=1$ Bi/cm (or, abA/cm)~\cite{LateralWaves-book}. The ambient medium is non-magnetic.

Next, we describe the requisite boundary conditions. Across the material sheet (at $z=0$) we impose~\cite{Bludov2013, margetis2016solutions}: (i) the continuity of the tangential component of the electric field; and (ii) a jump in the tangential component of the magnetic field that accounts for the surface current density, $\bold{j}^{s}$, induced by the tangential electric field on the sheet. These conditions are expressed by
\begin{subequations}\label{eqs:bcs-M}
\begin{gather}
\label{eq:bc2}
\left(\bold{E}_2-\bold{E}_1\right)\times\bold{e}_z=\bold{0},\\
\label{eq:bc1}
\left(\bold{H}_2-\bold{H}_1\right)\times\bold{e}_z=\frac{4\pi}{c}\bold{j}^{s}.
\end{gather}
\end{subequations}
Note that $\bold{j}^{s}$ macroscopically expresses the linear response of the conducting sheet and, thus, in principle includes hydrodynamic effects of the 2DES. In addition to the above conditions, the scalar components of the electromagnetic field $(\bold{E}_j, \bold{H}_j)$ must obey the Sommerfeld radiation condition as $\sqrt{x^2+y^2+z^2}\to\infty$ with $z\neq 0$~\cite{LateralWaves-book,margetis2016solutions}.

To describe the linear response of the conducting sheet, and thus express $\bold{j}^s$ in terms of the electric-field components parallel to the $xy$-plane, we follow the hydrodynamic approach for the 2D electron transport in the presence of a static magnetic field~\cite{kovtun2012lectures, fetter1985edge, cohen2018hall}.
Our model has two main ingredients, in the spirit of the Navier-Stokes equations: (i) the continuity  equation for
the electron number density; and (ii) the momentum equation, which accounts for forces acting on the 2DES as well as viscous effects.

Next, we describe the time-dependent, nonlinear Navier-Stokes-type equations for the electronic fluid, leaving aside the time-harmonic version of Maxwell's equations for a moment. Eventually, we linearize the hydrodynamic equations in the steady state, in correspondence to Eqs.~\eqref{eqs:curl-M} and~\eqref{eqs:bcs-M}.

Accordingly, the continuity equation is
\begin{subequations}
\begin{equation}
\label{eq:hydro1_0}
\partial_t n+\nabla\cdot (n\bold{v})=0,
\end{equation}
where $\partial_t=\partial/\partial t$ is the time derivative, $\nabla=(\partial/\partial x, \partial/\partial y)$ denotes the gradient on the $xy$-plane, $n(x,y,t)$ in the number (carrier) density, and $\bold{v}(x,y,t)$ is the velocity field for the electron flow. The momentum equation is
\begin{align}\label{eq:hydro2_0}
\partial_t\bold{v}+\bold{v}\cdot\nabla\bold{v}&=-\frac{\nabla P(n)}{mn}-\gamma\bold{v}+\eta\nabla\cdot\left(\nabla\bold{v}+\nabla\bold{v}^T-\nabla\cdot\bold{v}\operatorname{I}\right) \nonumber \\
+&\zeta \nabla\left(\nabla\cdot \bold{v}\right)-\left(\omega_c-\eta_H\nabla^2\right)\bold{v}\times\bold{e}_z-\frac{e\bold{E}_\parallel}{m}.
\end{align}
\end{subequations}
In the above, $\bold{E}_\parallel=(E_x,E_y)$ is the (time-dependent but not necessarily time-harmonic) tangential electric field on the sheet, at $z=0$. Since the electric field parallel to the $xy$-plane is continuous across the sheet, we do not need to specify the region index ($j=1,\,2$) for $\bold{E}_\parallel$, at $z=0$. We now define the various parameters appearing in the above equation as follows.
First, $\gamma$ is the scattering rate accounting for momentum dissipation due to the collisions of electrons with impurities and phonons; $m$ is the effective electron mass and $P$ is the internal pressure in the absence of interactions~\cite{fetter1985edge}; $-e$ is the (negative) electron charge; and $\omega_c = eB_{\text{st}}/\left(mc\right)$ is the cyclotron frequency associated with an externally applied, $z$-directed static magnetic field of strength $B_{\text{st}}$. Note that the term -$e\bold{E}_\parallel/m$ on the right-hand side of Eq.~\eqref{eq:hydro2_0} can be considered as the forcing that generates the electron flow, which amounts to nonzero $n$ and $\bold{v}$ on the conducting sheet. Of course, this forcing term is part of the solution of the full, coupled system that should include the time-dependent Maxwell equations (not written here in full generality).

The  viscous terms of Eq.~\eqref{eq:hydro2_0} deserve particular attention. These two distinct contributions involve the shear viscosity, $\eta$, and the Hall viscosity, $\eta_H$, which pertain to the effect of the diagonal and off-diagonal components of the electron viscosity tensor, respectively~\cite{cohen2018hall,alekseev2016negative,pellegrino2017nonlocal,avron1998}. In this setting, we also include the bulk viscosity, $\zeta$; see, e.g.,~\cite{cohen2018hall}. However, the bulk viscosity in graphene was shown to vanish (at least at high frequencies)~\cite{principi2016bulk, narozhny2019electronic}. In the following sections, we keep the contribution of the bulk viscosity, $\zeta$, in our analytical results, but neglect this contribution in our numerical calculations. Here, we assume that the material parameters of the 2DES are local, spatially homogeneous and time independent. (For results regarding $\omega$-dependent viscosities in the frequency domain, see Secs.~\ref{sec:poles} and~\ref{sec:amplitudes}.)

For a quantitative description of the parameters $\eta$ and $\eta_H$, particularly their dependence on $B_{\text{st}}$, we introduce the characteristic magnetic field $B_0=\hbar v_F k_F/(8 e \eta_0)$. This is expressed in terms of the Fermi wavenumber, $k_F$, the Fermi velocity, $v_F$, the electronic viscosity, $\eta_0$, at zero static magnetic field~\cite{principi2016bulk}, and the reduced Planck constant, $\hbar$. Accordingly, $\eta$ is equal to~\cite{berdyugin2019measuring, pellegrino2017nonlocal, principi2016bulk}
\begin{align*}
\eta=\frac{\eta_0}{1+\left(B_{\text{st}}/B_0\right)^2}.
\end{align*}
This formula is compatible with the definition of $\eta_0$ outlined above, since $\eta=\eta_0$ if $B_{\text{st}}=0$.
Furthermore, the Hall viscosity is given by~\cite{berdyugin2019measuring, pellegrino2017nonlocal, principi2016bulk}
\begin{align*}
\eta_H=-\eta_0 \frac{B_{\text{st}}/B_0}{1+\left(B_{\text{st}}/B_0\right)^2}.
\end{align*}
Note that $\eta_H\to 0$ as $B_{\text{st}}\to 0$, in contrast to the respective behavior of $\eta$. The Hall viscosity, $\eta_H$, seriously affects the dispersion of the 2D hydrodynamic waves in a static magnetic field \cite{alekseev2016negative, berdyugin2019measuring, pellegrino2017nonlocal}. Our sign convention for $\eta_H$ follows that in~\cite{pellegrino2017nonlocal}.

In correspondence to Eqs.~\eqref{eqs:curl-M} and~\eqref{eqs:bcs-M}, we linearize fluid equations~\eqref{eq:hydro1_0} and \eqref{eq:hydro2_0} around the equilibrium electron number density, $n_0$, and zero velocity field for the time-harmonic case. Hence, the resulting, time-harmonic electron flow excited by the dipole source is treated as a small perturbation of the static fluid that has electron density $n_0$. This flow gives rise to the surface current density $\bold{j}^s$; cf. Eq.~\eqref{eq:bc1}. Abusing notation, we use the same symbols for the time-harmonic dependent variables such as the fluid velocity ($\bold{v}$) and number density fluctuation ($n$) around $n_0$.

The linearized fluid equations in the $xy$-plane become
\begin{subequations}\label{eqs:fluid-th}
\begin{equation}
\label{eq:hydro1}
-\im\omega n+n_0\nabla\cdot \bold{v}=0,
\end{equation}
\begin{align}\label{eq:hydro2}
-\im\omega\bold{v}&=-\frac{s^2\nabla n}{n_0}-\left(\gamma-\eta\nabla^2\right)\bold{v}+\zeta\nabla\left(\nabla\cdot\bold{v}\right)\nonumber\\
& -\left(\omega_c-\eta_H\nabla^2\right)\bold{v}\times\bold{e}_z-\frac{e\bold{E}_\parallel}{m}.
\end{align}
\end{subequations}
Here, $s^2=m^{-1}(dP/d n)$ is the speed of the compressional waves at the equilibrium number density $n_0$. Equations~\eqref{eqs:fluid-th} are coupled with the Maxwell system of Eqs.~\eqref{eqs:curl-M} and~\eqref{eqs:bcs-M} in the three-dimensional space (see Fig.~\ref{fig0}).

Once the fluid velocity, $\bold{v}$, is determined as a linear function of the time-harmonic field $\bold{E}_\parallel$ in the 2DES from the coupled Eqs.~\eqref{eq:hydro1} and~\eqref{eq:hydro2}, the surface current density can be computed by $\bold{j}^s=-e n_0 \bold{v}$, where $en_0$ is the absolute value of the equilibrium electron charge density. This task is carried out in Sec.~\ref{sec:conductivity}.

\section{Nonlocal surface conductivity}
\label{sec:conductivity}

In this section, we develop the main ingredient of the linear-response theory for the hydrodynamic model of Sec.~\ref{sec:model}. More precisely, we derive the surface conductivity tensor of the infinite, translation invariant sheet by treating the time-harmonic electromagnetic field as the forcing in the  hydrodynamic equations. Hence, the emerging conductivity tensor takes into account the effects of shear and Hall viscosities of the 2DES, electronic compressibility and externally applied static magnetic field. Accordingly, this conductivity is spatially nonlocal (albeit translation invariant) and obeys the Onsager reciprocity relations~\cite{Onsager1931,Ziman-book}. We expect that the matrix elements of this emergent conductivity tensor can be obtained experimentally via nonlocal transport measurements; and may be used to estimate the values of the shear and Hall viscosity coefficients.

The procedure for obtaining the conductivity relies on solving Eqs.~\eqref{eq:hydro1} and~\eqref{eq:hydro2} for fixed tangential electric field, $\bold{E}_\parallel$ on the sheet. Thus, we can express the electron number density fluctuation, $n$, and the velocity, $\bold{v}$, as linear functions of $\bold{E}_\parallel$. The surface current density on the sheet is defined as
\begin{align*}
\bold{j}^s=-en_0 \bold{v}.
\end{align*}
We will now show that $\bold{j}^s$ and $\bold{E}_\parallel$ on the sheet (at $z=0$) are related by a convolution equation, viz.,
\begin{align*}
\bold{j}^s(\br)=\iint d\br'\ \underline{\sigma}(\br-\br';\omega)\,\bold{E}_\parallel(\br').
\end{align*}
Here, $\underline{\sigma}(\br;\omega)$ is a matrix valued kernel that corresponds to the surface conductivity tensor, whose Fourier representation is derived below; $\br=(x,y)$, and the domain of integration for the double integral is the
whole $xy$-plane.

Because the present setting is translation invariant in the $x$ and $y$ coordinates, we use the Fourier transform with respect to the position vector $\br=(x, y)$. We start with the application of this transform to Eqs.~\eqref{eq:hydro1} and~\eqref{eq:hydro2}. In this vein, consider the integral representations
\begin{gather*}
n(\br)=\frac{1}{4\pi^2}\iint d\bk\ \hat{n}(\bk)\,e^{\im\bk\cdot \br},\\
\bold{v}(\br)=\frac{1}{4\pi^2}\iint d\bk\ \hat{\bold{v}}(\bk)\,e^{\im \bk\cdot\br},\\
\bold{E}(\br,z)=\frac{1}{4\pi^2}\iint d\bk\ \hat{\bold{E}}(\bk,z)\,e^{\im\bk\cdot \br},
\end{gather*}
where $\bk=(k_x,k_y)$ is the Fourier variable, or in-plane wave vector, and the integration is carried out in the entire $k_x k_y$-plane ($-\infty< k_l<\infty$, $l=x,\,y$). In the above, $\hat{n}$, $\hat{\bold{v}}$ and $\hat{\bold{E}}$ are the Fourier transforms with respect to $\br$ of $n$, $\bold{v}$ and $\bold{E}$, respectively, while $z$ is kept fixed. Note that $\hat{\bold{E}}_\parallel(\bk)$ will denote the Fourier transform in $\br$ of $\bold{E}_\parallel(\br)$, at $z=0$.

We use the Fourier transform of Eq.~\eqref{eq:hydro1} in order to express $\hat{n}$ in terms of $\bk\cdot \hat{\bold{v}}$. Then, we substitute the result into the transformed Eq.~\eqref{eq:hydro2}. Consequently, we find a linear, nonhomogeneous system of equations for the components, $\hat{v}_x$ and $\hat{v}_y$, of $\hat{\bold{v}}$ where the wave vector $\bk$ enters as a parameter. By solving this system, we obtain formulas of the form $\hat{\bold{v}}(\bk)=\underline{\mathfrak G}(\bk;\omega) \hat{\bold{E}}_\parallel(\bk)$ where $\underline{\mathfrak G}$ is a $2\times 2$ matrix; its entries are computed explicitly but are not displayed here. By the formula $\hat{\bold{j}}^s=-en_0 \hat{\bold{v}}$, we thus find a relation of the form
\begin{align*}
\hat{\bold{j}}^s(\bk)=\underline{\hat\sigma}(\bk;\omega)\,\hat{\bold{E}}_\parallel(\bk),
\end{align*}
where the $2\times 2$ matrix $\underline{\hat\sigma}=[\hat{\sigma}_{ll'}]=-en_0 \underline{\mathfrak G}$ represents the (linear) surface conductivity tensor ($l,\,l'=x,\,y$). The entries for this $\underline{\hat\sigma}$, as functions of $\bk=(k_x,k_y)$ and $\omega$, are
\begin{subequations}\label{eqs:sigma-tensor}
\begin{gather}
\label{eq:sigma-xx}
\hat{\sigma}_{xx}=\frac{\im e^2 n_0}{m}\frac{\Omega(k)-\left(\frac{s^2}{\omega}-\im\zeta\right)k_y^2}{\Omega(k)\left(\Omega(k)-\left(\frac{s^2}{\omega}-\im\zeta\right)k^2\right)-\Omega^2_c(k)},\\
\label{eq:sigma-xy}
\hat{\sigma}_{xy}=\frac{\im e^2 n_0}{m}\frac{-\im\Omega_c(k)+\left(\frac{s^2}{\omega}-\im\zeta\right)k_xk_y}{\Omega(k)\left(\Omega(k)-\left(\frac{s^2}{\omega}-\im\zeta\right)k^2\right)-\Omega^2_c(k)},\\
\label{eq:sigma-yx}
\hat{\sigma}_{yx}=\frac{\im e^2 n_0}{m}\frac{\im\Omega_c(k)+\left(\frac{s^2}{\omega}-\im\zeta\right)k_xk_y}{\Omega(k)\left(\Omega(k)-\left(\frac{s^2}{\omega}-\im\zeta\right)k^2\right)-\Omega^2_c(k)},\\
\label{eq:sigma-yy}
\hat{\sigma}_{yy}=\frac{\im e^2 n_0}{m}\frac{\Omega(k)-\left(\frac{s^2}{\omega}-\im\zeta\right)k_x^2}{\Omega(k)\left(\Omega(k)-\left(\frac{s^2}{\omega}-\im\zeta\right)k^2\right)-\Omega^2_c(k)},
\end{gather}
\end{subequations}
where $k^2=k_x^2+k_y^2$, $\Omega(k)=\omega+\im\gamma+\im\eta k^2$, and $\Omega_c(k)=\omega_c+\im\eta_H k^2$.
Evidently, these formulas are Onsager-reciprocal~\cite{Onsager1931,Ziman-book}, viz.,  $\sigma_{xy}(B_{\text{st}})=\sigma_{yx}(-B_{\text{st}})$.

A few remarks on these results are in order. First, for $k = 0$ and $B_{\text{st}}=0$, each conductivity matrix element in
Eqs.~\eqref{eq:sigma-xx}--\eqref{eq:sigma-yy} has a Drude-like form, where the role of the Drude weight is played by the parameter $e^2n_0/m$. Second, we repeat at the risk of redundancy that the elements of the conductivity tensor can be experimentally obtained by nonlocal transport measurements~\cite{abanin2011giant, taychatanapat2013electrically, lundeberg2017tuning}. Thus, Eqs.~\eqref{eqs:sigma-tensor} can provide a reference model for future experimental investigations of surface conductivity on graphene; and help one estimate the value of the shear viscosity coefficient, $\eta$, by comparison to available experimental data.


It is worthwhile to compare our findings to previous works in the derivation of the surface conductivity tensor for the hydrodynamic regime of the 2DES. For example, similar results are obtained in~\cite{briskot2015collision} through the microscopic Boltzmann equation for the electron collisions. Here, Eq.~\eqref{eqs:sigma-tensor} \emph{additionally} includes the effects of the Hall viscosity, $\eta_H$, and a static magnetic field via the cyclotron frequency $\omega_c$. However, our starting point is different from that in~\cite{briskot2015collision}, since we rely on the (macroscopic) Navier-Stokes-type description.

In a similar vein, we should mention the results presented in~\cite{lovat2013semiclassical} for the intraband conductivity tensor of graphene, which are derived from the semiclassical Boltzmann equation. In particular, for sufficiently small $k$,
our formulas~\eqref{eqs:sigma-tensor} are in qualitative agreement with the conductivity derived in~\cite{lovat2013semiclassical}.

\section{Electric field: Integral Formulas}
\label{sec:integrals}

In this section, we obtain Fourier-Bessel integral representations for the components of the electric field generated by a vertical electric dipole, in the spirit of~\cite{margetis2016solutions,LateralWaves-book}. The radiating dipole is located at height $z_0$ above the 2D material ($z>0$); see Fig.~\ref{fig0}.
For this purpose, we solve Maxwell's equations for $(\bold{E}_j, \bold{H}_j)$ by using the nonlocal conductivity of Sec.~\ref{sec:conductivity} for the surface current density, $\bold{j}^s$, which enters the transmission boundary conditions across the sheet. An alternate approach, not followed here, is to solve directly the whole system of Eqs.~\eqref{eqs:curl-M}, \eqref{eqs:bcs-M} and~\eqref{eqs:fluid-th}, thus circumventing the use of the effective conductivity tensor.

We start by applying the Fourier transform with respect to $\br=(x,y)$ to Maxwell's equations. In this context, we need to write (cf. integral formulas of Sec.~\ref{sec:conductivity})
\begin{equation*}
\bold{H}\left(\br,z\right)=\frac{1}{4\pi^2}\iint d\bk\ \hat{\bold{H}}(\bk,z)\,e^{\im \bk\cdot \br},
\end{equation*}
where $\hat{\bold{H}}$ denotes the Fourier transform of the magnetic field, $\bk=(k_x,k_y)$ and the integration (in $\bk$) is performed in the entire $k_xk_y$-plane. The ensuing procedure consists of the following steps.
First, we solve the transformed Eqs.~\eqref{eq:m1} and~\eqref{eq:m2} for $\hat{\bold{E}}_j(\bk,z)$ and $\hat{\bold{H}}_j(\bk,z)$ in order to obtain $\hat{\bold{E}}_j(\bk,z)$ for $z\neq 0$.
Second, we need to determine the requisite integration constants. Therefore,  we have to use
boundary conditions~\eqref{eq:bc2} and~\eqref{eq:bc1}, at $z=0$. These conditions dictate the continuity of the tangential electric field and a jump of the tangential magnetic field across the sheet; the latter involves the surface current density, $\bold{j}^s(\br)$.
Accordingly, we express $\hat{\bold{j}}^s(\bk)$ in terms of $\hat{\bold{E}}_\parallel(\bk)=\hat{\bold{E}}(\bk,0)-\left(\hat{\bold{E}}(\bk,0)\cdot \bold{e}_z\right)\bold{e}_z $ by invoking the surface conductivity tensor described in
Eq.~\eqref{eqs:sigma-tensor}. Details of this procedure can be found in Appendix~\ref{sec:derivation_E}. Note that the component $H_z$ vanishes identically in our problem.

Because of the axisymmetry of our geometry (Fig.~\ref{fig0}), it is convenient to use the cylindrical coordinates
$(r,\phi,z)$ where $r=\sqrt{x^2+y^2}$ and $0\le\phi< 2\pi$; $x=r\,\cos\phi$ and $y=r\,\sin\phi$. This choice enables us to convert the aforementioned Fourier representations for the electromagnetic field to one-dimensional (Fourier-Bessel) integrals with respect to the polar coordinate $k$ of the Fourier space~\cite{LateralWaves-book}. After some algebra, for a \emph{unit} Hertzian electric dipole we obtain the following integrals for the electric field:
\begin{subequations}\label{eqs:E-FB}
\begin{gather}
\label{eq:er}
E_r(r,z)=\frac{\im}{\omega\e}
\begin{cases}
\begin{split}
&\int_0^{\infty}dk\, k^2J_1\left(kr\right)\left[\frac{\mathcal A(k) + \mathcal D(k)}{\mathcal D(k)} e^{-\beta(k)\left(z+z_0\right)}\right.\\
&\qquad  \left.+\sgn(z-z_0)e^{-\beta(k)\left|z-z_0\right|}\right], \enspace z\ge 0,
\end{split}
\\
{\displaystyle \int_0^{\infty}dk\,k^2 J_1\left(kr\right)\frac{\mathcal A(k)}{\mathcal D(k)}e^{\beta(k)\left(z-z_0\right)}}, \enspace z\le 0;
\end{cases}\\
\label{eq:efi}
E_{\phi}(r,z)=-\frac{\omega^2 D_0}{c^2 \e}
\begin{cases}
{\displaystyle \int_0^{\infty}dk \,k^2 J_1\left(kr\right)\frac{\Omega_c(k)}{\mathcal D(k)}e^{-\beta(k)\left(z+z_0\right)}}, & z\ge 0,
\\
{\displaystyle\int_0^{\infty}dk\,k^2 J_1\left(kr\right)\frac{\Omega_c(k)}{\mathcal D(k)}e^{\beta(k)\left(z-z_0\right)}}, & z\le 0;
\end{cases}\\
\label{eq:z}
E_z(r,z)=\frac{\im}{\omega\e}
\begin{cases}
\begin{split}
&\int_0^{\infty}dk\, \frac{k^3}{\beta(k)} J_1\left(kr\right)\left[\frac{\mathcal A(k) + \mathcal D(k)}{\mathcal D(k)}e^{-\beta(k)\left(z+z_0\right)}\right.\\
&\qquad \left.+e^{-\beta(k)\left|z-z_0\right|}\right], \enspace z> 0,\quad (r, z)\neq (0, z_0),
\end{split}
\\
{\displaystyle \int_0^{\infty}dk \,k^3 J_1\left(kr\right)\frac{\mathcal A(k)}{\mathcal D(k)\beta(k)} e^{\beta(k)\left(z-z_0\right)}}, \enspace z< 0.
\end{cases}
\end{gather}
\end{subequations}
Here, $\sgn(z)$ is the signum function, $\sgn(z)=\pm 1$ if
$\pm z>0$; $J_1$ is the Bessel function of the first order; and the electric moment, $I_0\ell$, of the dipole is set equal to unity (cf. Sec.~\ref{sec:model}). We also define the following quantities:
\begin{gather*}
\begin{split}
\mathcal A(k)&=\left\{\left(s^2-\im\omega\zeta\right) k^2-\omega\Omega(k)\right\}\left(k^2D_0 +\e\omega\Omega(k)\beta(k)\right)\\
&+\omega^2\e\beta(k) \Omega^2_c(k),
\end{split}\\
\begin{split}
\mathcal D(k)=&-\mathcal A(k)
-\beta(k) k^2D_0^2/\e-D_0 \omega\Omega(k)\beta^2(k),
\end{split}
\end{gather*}
and $\beta(k) = \sqrt{k^2 - k_0^2}$. Note that
$k_0=\omega\sqrt{\e}/c$ is the wavenumber of the ambient dielectric medium, and $D_0=2\pi e^2n_0/m$ expresses the Drude weight. In addition, we require that the integration in the $k$ variable is carried out under the condition
\begin{align*}
\Re\beta(k)>0,
\end{align*}
which ensures that the electromagnetic field decays as $|z|\to \infty$. Recall that $k=|\bk|=\sqrt{k_x^2+k_y^2}$.

A few remarks on Eq.~\eqref{eqs:E-FB} are in order.
We observe that the angular component, $E_\phi$, of the electric field vanishes identically in the absence of a static magnetic field, when $B_{\text{st}}=0$ and, thus, $\omega_c=0$ and $\eta_H=0$. This simplified electric-field polarization is in agreement with previous studies in dipoles radiating over the isotropic and homogeneous graphene, in the ohmic regime~\cite{margetis2016solutions}. Interestingly, the $z$-directed static magnetic field, $B_{\text{st}}\bold{e}_z$, applied to the 2DES is solely responsible for the generation of the (time-harmonic) $\phi$-component of the electric field here, through the Hall viscosity, $\eta_H$, and cyclotron frequency, $\omega_c$. In Sec.\ref{sec:amplitudes}, we will show that two collective modes, both a diffusive mode and a plasmon, can manifest in this angular component; and, thus, can exhibit a non-longitudinal character.

We close this section with a more technical remark. Because the components $E_r$ and $E_\phi$ are continuous across $z=0$, the $z$ coordinate can be set equal to $0$ in their formulas for each region ($z>0$ or $z<0$) without ambiguity. In contrast, $E_z$ exhibits a jump proportional to the surface charge density at $z=0$; thus, its formula for each region yields a different limiting value as $z$ approaches $0$.

\section{Singularities and collective modes}
\label{sec:poles}

In this section, we focus on the role of singularities that are present in the integrands of the Fourier-Bessel representation for the electric field (Sec.~\ref{sec:integrals}). There are two types of such singularities, namely, poles and branch points. These can admit distinct physical interpretations. In particular, some poles are associated to collective modes in the 2DES. We derive and discuss the relevant dispersion relations.

Consider the Fourier-Bessel integrals in Eq.~\eqref{eqs:E-FB}. We can view the Fourier variable, $k$, as complex, and examine the analytic continuation of each integrand in the complex $k$-plane. The singularities of the integrands as functions of $k$ are: (a) The branch points $k=\pm k_0$, which are due to the multivalued even function $\beta(k)=\sqrt{k^2-k_0^2}$; and (b) simple poles, which come exclusively from the (complex) zeros of the denominator
$\mathcal D(k)$ (cf. Eq.~\eqref{eqs:E-FB}). The poles can give rise to surface waves, or collective modes, on the sheet.

Recall that we require $\Re\beta(k)>0$ so that the corresponding scattered wave, which has wavenumber $i\beta$  in the $z$-direction, decays with the height $|z|$ (Sec.~\ref{sec:integrals}). This condition on $\beta(k)$ defines the physical branch (or ``top Riemann sheet'') of the function $\beta(k)$. For fixed $\omega$, a zero $k=k_*(\omega)$ of $\mathcal D(k)$ is considered  \emph{physically admissible} if it obeys the following conditions:
\begin{align}\label{eq:cond-k}
\Im k_*(\omega)\ge 0\quad \mbox{and}\quad \Re\beta\left(k_*(\omega)\right)>0.
\end{align}
The first condition implies that the amplitude of the respective surface wave mode, which comes from the residue at the pole $k=k_*$ of the Fourier-Bessel transform, does not grow with the radial distance, $r$.

Next, we elaborate on the character of each singularity.

\paragraph{Branch points, $k=\pm k_0$.} The effect of these singularities is intimately connected to retardation in the ambient medium. By properly deforming the integration path for the field components in the upper complex $k$-plane for sufficiently large radial distance, $r$, we can split each Fourier-Bessel integral into distinct contributions. One of these terms comes from the infinite cut associated with $k_0$. This contribution is interpreted as the radiation field into the unbounded dielectric medium with radial wavenumber equal to $k_0$. For a lossless ambient medium ($k_0>0$), we expect that $k_0$  can be much smaller than the real parts of the wavenumbers of the surface modes, in a suitable range of frequencies.

\paragraph{Poles.} We now address the zeros $k_*(\omega)$ of the denominator $\mathcal D(k)$ in the integrands of Eq.~\eqref{eqs:E-FB}, for given $\omega$.
By setting $\mathcal D(k)$ equal to zero, we obtain the relation
\begin{align}\label{eq:dr}
&\left[\eta^2+\eta_H^2+ \frac{\im\eta\left(s^2-\im\omega\zeta\right)}{\omega}\right]\beta(k)^5+\frac{\im D_0\eta}{\e\omega}\beta(k)^4\nonumber\\
&+\left[2\eta\omega\left(\frac{\eta\omega}{c^2}+\frac{\gamma}{\omega}\right)+2\eta_H\omega\left(\frac{\eta_H\omega}{c^2}+\frac{\omega_c}{\omega}\right)\right.\nonumber\\
&\left. \qquad +s^2\left(1+2\im\frac{\eta\omega}{c^2}+\im\frac{\gamma}{\omega}\right)\right]\beta(k)^3\nonumber\\
&+\left[\omega^2\left(\frac{\eta\omega}{c^2}+\frac{\gamma}{\omega}-\im\right)\left(\frac{\eta\omega}{c^2}+\frac{\gamma}{\omega}-\im+\im\frac{s^2-\im\omega\zeta}{c^2}\right)\right.\nonumber\\
&\left. \qquad+\omega^2\left(\frac{\eta_H\omega}{c^2}+\frac{\omega_c}{\omega}\right)^2+\frac{D_0^2}{c^2\e^2}\right]\beta(k)\nonumber\\
&+\frac{D_0}{\e}\left(\im\frac{\gamma}{\omega}+1+\frac{s^2-\im\omega\zeta}{c^2}\right)\beta^2(k)\nonumber\\
&-\im\frac{D_0\omega^2}{\e c^2}\left(\frac{\eta\omega}{c^2}+\frac{\gamma}{\omega}-\im+\im \frac{s^2-\im\omega\zeta}{c^2}\right)=0.
\end{align}
Note that the Fourier variable, $k$, enters the above  relation only through $\beta(k)$. Evidently, the left-hand side of this equation is a fifth-degree polynomial in $\beta$; thus, it has exactly five complex roots $\beta(k_*)$. Accordingly, for any given frequency $\omega$, we seek the physically admissible solutions $k_*(\omega)$ of Eq.~\eqref{eq:dr} via the roots $\beta$. Thus, all zeros $k_*(\omega)$ can be grouped into five pairs of symmetric-through-the-origin points in the complex $k$-plane. By statement~\eqref{eq:cond-k}, we can admit at most one $k=k_*(\omega)$ from each pair.

Next, we derive approximate closed-form expressions for $k_*(\omega)$ in the nonretarded frequency regime, assuming~\cite{Bludov2013}
\begin{align*}
	\frac{D_0}{\omega c\e}\ll 1.
\end{align*}
We switch off the static magnetic field ($B_{\text{st}}=0$), and neglect the effects of internal pressure ($s=0$) and bulk viscosity ($\zeta=0$). Hence, Eq.~\eqref{eq:dr} is simplified. After some algebra, we write this relation as
\begin{align*}
&\left[\eta\beta^2+\omega\left(\frac{\eta\omega}{c^2}+\frac{\gamma}{\omega}-\im\right)\right]\ \left[\eta\beta^3+\omega\left(\frac{\eta\omega}{c^2}+\frac{\gamma}{\omega}-\im\right)\beta\right.\\
&\left. \quad +\im\frac{D_0}{\omega \e}\beta^2-\im\frac{D_0\omega}{\e c^2}\right]=0.
\end{align*}
Our task is to solve this equation subject to statement~\eqref{eq:cond-k}.

A physically admissible solution $k_*(\omega)=k_d(\omega)$ is
\begin{equation}
\label{eq:kd}
k_{d}(\omega)\simeq\sqrt{\frac{i\omega-\gamma}{\eta}+\frac{\omega^2}{c^2}\left(\e-1\right)},
\end{equation}
which is independent of $D_0$.
The respective surface wave is a \emph{diffusive mode}~\cite{ShearGregory, lucas2016sound}.
Notably, by dispersion relation~\eqref{eq:kd} this mode exhibits an appreciable decay. In particular, for $\gamma =0$ the damping ratio in vacuum ($\e=1$) equals $\Im k_d(\omega)/\Re k_d(\omega)=1$. Interestingly, for $\eta\omega/c^2\ll 1$ we have $k_{d}(\omega)\approx\sqrt{\frac{i\omega-\gamma}{\eta}}$ and $\Re k_d(\omega) \gg k_0(\omega)$, which indicates a subwavelength (albeit highly damped) diffusive mode.

Another admissible solution, $k_*(\omega)=k_{pl}(\omega)$, corresponds to a surface \emph{plasmon}. For sufficiently small shear viscosity, i.e., if $\eta\e\omega/c^2\ll 1$, we find
\begin{equation}
\label{eq:kpl}
k_{pl}(\omega)\simeq\frac{\omega\e\left(\omega+\im\gamma\right)}{D_0}\left(1-\im\eta\frac{\omega^2\left(\omega+\im\gamma\right)\e^2}{ D_0^2}\right).
\end{equation}
In the special case with $\eta=0$, this formula reduces to the familiar Drude-like dispersion relation, according to which $\omega \sim \sqrt{k}$~\cite{Bludov2013}. Equation~\eqref{eq:kpl} provides a small correction term due to the nonzero shear viscosity, where $\eta=\eta_0$ for zero static magnetic field, $B_{\text{st}}=0$.

So far, our explicit analytical results for the solution $k_*(\omega)$ have not taken into account the effect of the static magnetic field, $B_{\text{st}}$. Next, we consider the leading-order correction to Eq.~\eqref{eq:kd} for the diffusive mode caused by the Hall viscosity, $\eta_H$, because of a nonzero but weak field $B_{\text{st}}$.

To this end, we turn our attention to Eq.~\eqref{eq:dr}. We seek a solution $k=k_*(\omega)$ of this equation in the form of a perturbation expansion in powers of $B_{\text{st}}$, viz.,
\begin{equation*}
k_*\simeq k_d^{(0)}+k_d^{(1)}+k_d^{(2)}
\end{equation*}
where $k_d^{(0)}$ is given by Eq.~\eqref{eq:kd} while $k_d^{(j)}$ is proportional to
$B_{\text{st}}^j$ for $j=1,\,2$; here, we assume that $|\eta_H\omega/c^2|\ll 1$ and $\omega_c/\omega\ll 1$. Because of this choice for $k_d^{(0)}$, we can consider the above approximate solution for $k_*$ as an expansion for the diffusive mode. By direct substitution of this expansion for $k_*$ into Eq.~\eqref{eq:dr} and application of dominant balance in the small parameters $\eta_H\omega/c^2$ and $\omega_c/\omega$, we obtain
\begin{equation*}
\begin{split}
&k_d^{(1)}=0\quad \mbox{and}\quad \\
&k_d^{(2)}=-\sqrt{\frac{\frac{\im\omega-\gamma}{\eta}-\frac{\omega^2}{c^2}}{\frac{\im\omega-\gamma}{\eta}+\frac{\omega^2}{c^2}\left(\e-1\right)}}\frac{\left(\im\frac{\eta_H}{\eta}\left(\omega+\im\gamma\right)+\omega_c\right)^2}{6\im\omega\eta\sqrt{\frac{\im\omega-\gamma}{\eta}}-\frac{2D_0}{\omega\e}\left(\omega+\im\gamma\right)}.
\end{split}
\end{equation*}

In this regime, it is of interest to express the wavenumber of the diffusive mode in terms of the static magnetic field, $B_{\text{st}}$. By invoking the relevant formulas for the shear and Hall viscosities (Sec.~\ref{sec:model})~\cite{alekseev2016negative, berdyugin2019measuring, pellegrino2017nonlocal}, we obtain
\begin{equation*}
\begin{split}
&k_*(\omega)=k_d(\omega)\simeq\sqrt{\frac{i\omega-\gamma}{\eta}+\frac{\omega^2}{c^2}\left(\e-1\right)}\\
&-\left(\frac{B_{\text{st}}}{B_0}\right)^2\sqrt{\frac{\frac{\im\omega-\gamma}{\eta}-\frac{\omega^2}{c^2}}{\frac{\im\omega-\gamma}{\eta}+\frac{\omega^2}{c^2}\left(\e-1\right)}}\frac{\left(-\im\left(\omega+\im\gamma\right)+\frac{eB_0}{mc}\right)^2}{6\im\omega\eta\sqrt{\frac{\im\omega-\gamma}{\eta}}-\frac{2D_0}{\omega\e}\left(\omega+\im\gamma\right)}.
\end{split}
\end{equation*}
Notably, this dispersion relation does not depend on the direction  of the static magnetic field which is perpendicular to the material sheet. It is worthwhile to reiterate that the shear and Hall viscosities, $\eta$ and $\eta_H$, whose effect is captured in a perturbative sense here, cause a nonlocal linear response via the surface conductivity tensor; recall Eq.~\eqref{eqs:sigma-tensor}. To our knowledge, the above expansion for $k_d(\omega)$ forms an extension of previous results found in the literature~\cite{ShearGregory, lucas2016sound}.

We now proceed to study numerically the dispersion relations for the plasmon and diffusive modes, after gaining some insight from perturbation theory. We therefore compute the roots $k=k_*(\omega)$ of the equation $\mathcal D(k;\omega)=0$ numerically, for a range of THz frequencies. Accordingly, we plot the frequency $\omega$ versus the real and imaginary parts of the in principle complex wavenumber $k=k_*(\omega)$. Let $k_*=k_*'+\im k_*''$ where $k_*'$ and $k_*''$ are the real and imaginary parts of $k_*$, respectively; $k_*'=\Re k_*$ and $k_*''=\Im k_*$. In our plots, we distinguish the two modes, $k_*=k_{pl}$ or $k_*=k_{d}$.

We alert the reader that we refer to the ``local effect'' regarding the dispersion relations in situations with an underlying $\bk$-\emph{independent} surface conductivity tensor (Sec.~\ref{sec:conductivity}). Considering the minimal set of parameters that are responsible for the dependence of this conductivity on $\bk=(k_x,k_y)$, for locality to occur we set each of these parameters equal to zero. Thus, we set $\eta_0=0$ and $s^2=0$, while allowing for a nonzero static magnetic field, $B_{\text{st}}$; consequently, $\eta=0$ and $\eta_H=0$ regardless of the value of $B_{\text{st}}$. In contrast, the ``nonlocal effect'' regarding the dispersion relations arises when the conductivity tensor becomes strictly $\bk$-dependent.

Note that in our numerics we use two different models for the characteristic viscosity, $\eta_0$, at zero static magnetic field, when the nonlocal effect is present. According to one model for $\eta_0$, we take $\eta_0=\eta_0(\omega)$, a function of frequency which is derived in~\cite{pellegrino2017nonlocal}.
In another model, we use a constant $\eta_0$, i.e.,
$\eta_0=0.05\,\text{m}^2\text{s}^{-1}$ which is a value measured experimentally at zero frequency~\cite{bandurin2016negative, berdyugin2019measuring}.

\begin{figure*}[h]
\includegraphics[width=1\linewidth]{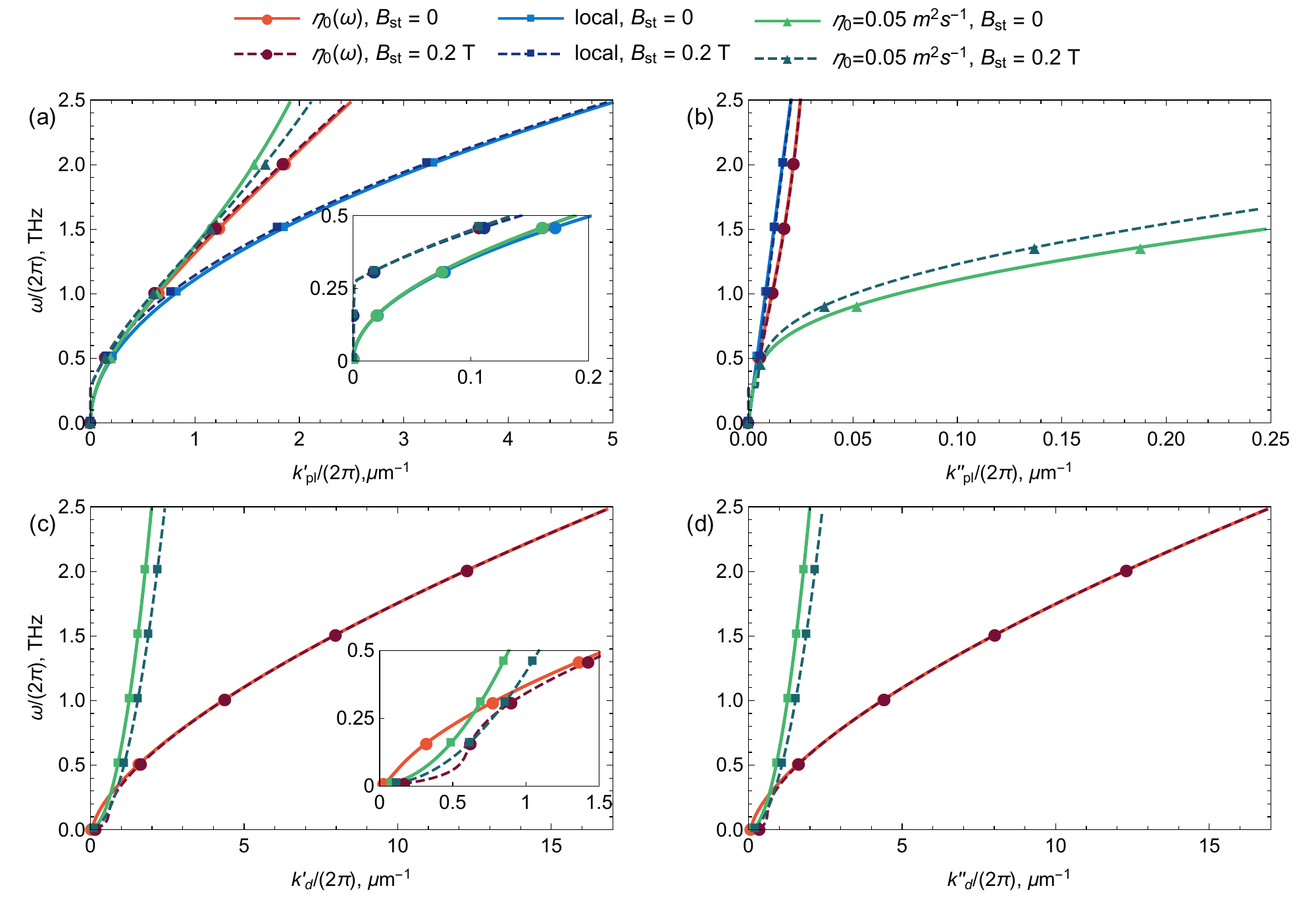}
\centering{}\caption{(Color online) Numerically computed dispersion relations for collective modes on a material sheet via solutions of Eq.~\eqref{eq:dr}. Top panels [(a), (b)]: Frequency $\omega/(2\pi)$ versus real part $k'_{pl}/(2\pi)$ [(a)] and imaginary part $k_{pl}''/(2\pi)$ [(b)] of wavenumber $k_{pl}/(2\pi)$ of the plasmon mode. Bottom panels [(c), (d)]: Frequency $\omega/(2\pi)$ versus real part $k'_{d}/(2\pi)$ [(c)] and imaginary part $k_{d}''/(2\pi)$ [(d)] part of wavenumber $k_{d}/(2\pi)$ of the diffusive mode. Values of the material parameters used in numerics for all plots are: $\gamma=0.01\,$THz and $n_0=10^{12}\,$cm$^{-2}$.  The value of the static magnetic field is $B_{\text{st}}=0$ or $0.2\,$T. The ``local'' effect amounts to $s=0$ and $\eta_0=0$ (blue solid and dashed curves in panels (a), (b)); otherwise, $s=0.7\times10^5$ m s$^{-1}$ and $\eta_0=\eta_0(\omega)$ by~\cite{pellegrino2017nonlocal} or $\eta_0=0.05\,$m$^2$s$^{-1}$~\cite{bandurin2016negative, berdyugin2019measuring}.}
\label{fig:disp_rel}
\end{figure*}

 Figure~\ref{fig:disp_rel} aims to demonstrate the relative influence of the nonlocal effect and static magnetic field on the dispersion relations of the plasmon and diffusive modes. Let us focus on the plasmon first (Figs.~\ref{fig:disp_rel}(a),(b)). Our numerics confirm that the dispersion curve $\omega(k'_{pl})$ of the plasmon (Fig.~\ref{fig:disp_rel}(a)) exhibits the familiar asymptotic behavior $\omega \sim \sqrt{k'_{pl}}$ for sufficiently large $k_{pl}'$ in the nonretarded frequency regime, if the nonlocal effect in the surface conductivity is switched off. Accordingly, when the local effect dominates, for nonzero magnetic field ($B_{\text{st}}=0.2\,$T in our numerics) we observe the expected gap in the relation $\omega(k'_{pl})$ at $k'_{pl}=0$, in agreement with previous results~\cite{AlekseevMagnetosound, Counterflow}; see the inset of Fig.~\ref{fig:disp_rel}(a). Now consider the nonlocal effect in this setting. By Fig.~\ref{fig:disp_rel}(a), the slope of the plasmon dispersion relation, $\omega(k'_{pl})$, then increases. Thus, the plasmon group velocity or speed $d\omega/d k_{pl}'$
 increases; compare the solid red and dashed blue curves in
 Figs.~\ref{fig:disp_rel}(a),(b). This trend is more pronounced at higher frequencies $\omega$ where the dominant nonlocal effect is associated with the compressional waves, i.e., the pressure term proportional to $s^2$ in the linearized momentum equation~\eqref{eq:hydro2} (Sec.~\ref{sec:model}). (A similar conclusion can be drawn in the case with nonzero bulk viscosity). This result is consistent with previous studies of hydrodynamic effects in 2D materials~\cite{giuliani2005quantum}. We repeat at the risk of redundancy that the viscosity, $\eta_0$, at zero static magnetic field can be a function of the frequency $\omega$~\cite{pellegrino2017nonlocal}. Therefore, in the plots of Fig.~\ref{fig:disp_rel} we consider both the cases with $\eta_0=\eta_0(\omega)$ and a constant $\eta_0$ which has the zero-frequency value $\eta_0\equiv 0.05\,$m$^2$/s~\cite{bandurin2016negative, berdyugin2019measuring}, for comparison purposes.

\onecolumngrid

\twocolumngrid

Next, we turn our attention to the dispersion relation, $k_d(\omega)=k_d'(\omega)+\im k_d''(\omega)$, of the diffusive mode, when the nonlocal effect is present ($s\neq 0$ and $\eta_0\neq 0$);
see Figs.~\ref{fig:disp_rel} (c), (d). In contrast to the plasmon mode, in the case of the diffusive mode there is no gap in the $\omega(k_d')$ if $B \ne 0$.  Notably, the real part, $k_d'=\Re k_d$, of the wavenumber, $k_d$, of the diffusion mode can be much larger than the real part, $k_{pl}'$, of the plasmon wavenumber, $k_{pl}$; cf. Figs.~\ref{fig:disp_rel}(a), (c). Thus, the diffusive mode can provide even \emph{higher confinement of the electromagnetic radiation} than the plasmon at THz frequencies. On the other hand, the imaginary part, $k_d''=\Im k_d$, of the diffusive mode wavenumber is at least as large as the real part, $k_d'$. Hence, as we point out in Sec.~\ref{sec:poles}, the diffusive mode, unlike the conventional plasmon in graphene, can experience high dissipation in the 2D material.

\begin{figure*}[h]
\includegraphics[width=1\linewidth]{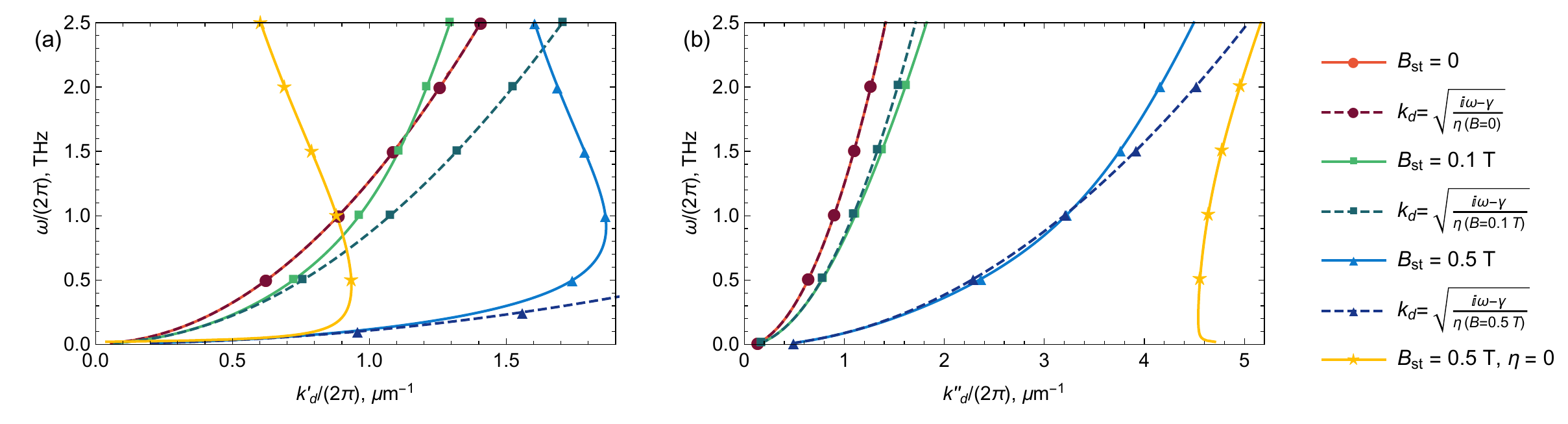}
\centering{}\caption{(Color online)
Numerically computed frequency $\omega/(2\pi)$ versus real part $k'_{d}/(2\pi)$ [(a)] and imaginary part $k_{d}''/(2\pi)$ [(b)] of wavenumber $k_{d}/(2\pi)$ of the diffusive mode by Eq.~\eqref{eq:dr}; and respective analytical result from Eq.~\eqref{eq:kd}. Values of parameters used in numerics are: $B_{\text{st}}=0,\,0.1,\,0.5\,$ T; $\gamma=0.01\,$THz, $n_0=10^{12}\,$cm$^{-2}$, $\eta_0 = 0.1\,$ m$^2$s$^{-1}$, and $s=0.7\times10^5\,$m s$^{-1}$. For comparison purposes, we also use the test value $\eta=0$ with $B_{\text{st}}=0.5\,$T (solid yellow curves).
 A negative group velocity is evident at frequencies $\omega/(2\pi)>1$ THz.}
\label{fig_diff}
\end{figure*}

Figure~\ref{fig_diff} depicts the numerically computed dispersion relation for the diffusive mode by Eq.~\eqref{eq:dr} as well as analytical formula~\eqref{eq:kd}. Interestingly, the diffusive mode can exist even when the shear viscosity is zero ($\eta \equiv 0$) but a nonzero static magnetic field is applied ($B_{\text{st}} \ne 0$). In this situation, the momentum transport within the electron fluid is characterized by the Hall viscosity, $\eta_H$. The dispersion relation of the resulting Hall diffusive mode is shown in Fig. \ref{fig_diff}; see the yellow curve for which we set $\eta \equiv 0$ for illustrative purposes. When both viscosities $\eta$ and $\eta_H$ are taken into account ($\eta\neq 0$ and $\eta_H\neq 0$) and the static magnetic field is present, the effective diffusive mode combines the features of the conventional~\cite{ShearGregory, lucas2016sound} and Hall diffusions. In fact, in Fig.~\ref{fig_diff}, compare the data of the green (for $B=0.1\,$T) and blue (for $B =0.5\,$T) curves to the cases of conventional diffusion~\cite{ShearGregory, lucas2016sound} (red curve, $B=0$) and the Hall diffusion mode (yellow curve). Interestingly, in this regime, the group velocity of the diffusive mode becomes negative at frequencies $\omega/(2\pi)>1$ THz.

\onecolumngrid

\twocolumngrid

We conclude this section by reiterating that so far we derived dispersion relations for two distinct modes on the material sheet. A pending question is whether these modes can actually be excited by a radiating, electric-current-carrying source. Aspects of this question, particularly
the amplitudes of the modes and their comparisons to the accompanying radiation field, are investigated in Sec.~\ref{sec:amplitudes}.

\section{Excitation of hydrodynamic modes}
\label{sec:amplitudes}

In this section, we numerically evaluate the integrals describing the electric field components on the conducting sheet, when the current-carrying source is a vertical Hertzian electric dipole (see Fig.~\ref{fig0}). By virtue of this computation, we discuss the excitation and relative contributions of the plasmon and diffusion modes. As pointed out in Sec.~\ref{sec:poles}, the surface waves due to these modes co-exist with the radiation field which has a complicated structure. In this vein, we study the possible dominance of the a single mode in the near- or   intermediate-field region of the source.

\subsection{Methodology}
\label{subsec:method}

We start with the Fourier-Bessel integral representations of Eq.~\eqref{eqs:E-FB}. Here, we choose to focus on the computation of the radial component, $E_r$, and angular component, $E_\phi$, of the electric field for illustrative purposes. These components form the electric field parallel to the sheet, and thus are continuous across $z=0$ for an elevated dipole at height $z_0$ ($z_0>0$). Since all fields decay away from the boundary, in our numerical evaluation we let $z\to 0$ and $z_0\to 0$ with $z<z_0$. In other words, both the Hertzian dipole (from region 1) and the observation point are allowed to approach the sheet. The requisite Fourier-Bessel integrals become
\begin{subequations}\label{eqs:int-rphi}
\begin{gather}
\label{int_r}
E_r(r,0)=\frac{i}{\omega\e}\int_0^{\infty}dk\ k^2 J_1(kr)\frac{\mathcal A(k)}{\mathcal D(k)},\\
\label{int_phi}
E_{\phi}(r,0)=-\frac{\omega^2 D_0}{c^2\e}\int_0^{\infty}dk\ k^2J_1(kr)\frac{\omega_c+\eta_H k^2}{\mathcal D(k)},
\end{gather}
\end{subequations}
where we have set the dipole electric moment, $I_0\ell$, equal to unity (Sec.~\ref{sec:model}). Recall that the functions $\mathcal A(k)$ and $\mathcal D(k)$ are defined in Sec.~\ref{sec:integrals}. Note that the $z$-directed component of the electric field, $E_z$, is qualitatively similar to the radial one, $E_r$. Indeed, one can show that Eq.~\eqref{eq:z} can be approximated by Eq.~\eqref{eq:er} (up to the singularities due to the dipole) in the $k$-region of interest where $k\gg k_0$ and, hence, $k/\beta\approx 1$.

We proceed to numerically evaluate integrals~\eqref{int_r} and~\eqref{int_phi} by use of two different techniques. First, we use the ``integration then summation'' technique \cite{blakemore1976comparison, lucas1995evaluating} which is accelerated by the ``$\epsilon$-algorithm''~\cite{wynn1956device}. This numerical method of ``integration then summation'' is designed for integration of functions containing oscillatory terms such as the Bessel function $J_1(\wp)$ for real $\wp$. Second, we apply notions of contour integration in the complex plane, particularly the residue theorem, to confirm the accuracy of the above numerical integration. To this end, we calculate the sum of the contributions to each Fourier-Bessel integral from: (i) the cut originating from the branch point $k=k_0$ in the complex $k$-plane (see Sec.~\ref{sec:poles});  and (ii) the admissible poles of the integrands, which pertain to the plasmon and diffusive modes. Abusing terminology slightly, we often refer to the contribution to integration from the cut (related to $k=k_0$) as the branch point contribution. It should be borne in mind that this loose statement becomes reasonably accurate only asymptotically, for sufficiently large $k_0 r$, in the far-field region.

Next, we provide some details of our methodology for the separation of branch cut and residue contributions. To facilitate manipulations in the complex $k$-plane, we first write the Bessel function $J_1$ as $J_1(\wp)=[H^{(1)}(\wp)+H^{(2)}(\wp)]/2$, in terms of the first- and second-kind Hankel functions $H_1^{(j)}(\wp)$ ($j=1,\,2$), of which only $H_1^{(1)}(\wp)$ satisfies the acceptable radiation condition as $r\to\infty$; $\wp=k r$. Subsequently, we express each integrand in terms of $H_1^{(1)}(kr)$ only, by using the symmetry property $H_1^{(2)}(e^{-\im \pi}\wp)=H_1(\wp)$ with $\wp=kr$, and extending the integration path in the $k$ variable over the whole real axis ($-\infty < k<+\infty$). Furthermore, regarding the multivalued function $\beta(k)=(k^2-k_0^2)^{1/2}$, we choose the cuts emanating from $\pm k_0$ to be straight half lines parallel to the positive and negative real axes in the complex $k$-plane.

To single out the branch cut contribution, we deform the integration path for $E_r$ and $E_\phi$ from the real axis to a contour that is wrapped around the cut in the upper half $k$-plane, through a large semicircle. Along the cut, set $k=k_0(1+\im\tau)$ where $\tau>0$.
The corresponding branch point contributions to the Fourier-Bessel integrals can be written as
\begin{gather*}
\begin{split}
E_r^{bp}(r,0)=\frac{k_0^3}{2\omega\e}&\int_0^{\infty}d\tau\ \left(1+\im\tau\right)^2H_1^{(1)}\bigl(k_0r(1+\im\tau)\bigr)\\
&\times\left[\frac{\mathcal A^-(k_0(1+\im\tau))}{\mathcal D^-(k_0(1+\im\tau))}-\frac{\mathcal A^+(k_0(1+\im\tau))}{\mathcal D^+(k_0(1+\im\tau))}\right],
\end{split}\\
\begin{split}
&E_{\phi}^{bp}(r,0)=\frac{\im\omega^2 D_0k_0^3}{2c^2\e}\int_0^{\infty}d\tau\ \left(1+\im\tau\right)^2H^{(1)}_1\left(k_0r\left(1+\im\tau\right)\right)\\
&\times\left(\omega_c+\eta_Hk_0^2\left(1+\im\tau\right)^2\right) \left[\frac{1}{\mathcal D^-(k_0(1+\im\tau))}-\frac{1}{\mathcal D^+(k_0(1+\im\tau))}\right].
\end{split}
\end{gather*}
Here, the symbol $\mathcal X^{\pm}$ (for $\mathcal X=\mathcal A, \mathcal D$) denotes the value of $\mathcal X(k)$ on the right ($+$) or left ($-$) side of the upper cut where $\beta(k)=\pm k_0 e^{i\pi/4}\sqrt{2+i\tau}$. Note that the value of the function $\sqrt{2+\im \tau}$ at $\tau=0$ is $\sqrt{2}$; more generally, $\Re\sqrt{2+\im\tau}>0$ for all $\tau\ge 0$. In the asymptotic regime with $k_0 r\gg 1$, the major contribution to integration in the integrals for $E_{r}^{bp}$ and $E_{\phi}^{bp}$ comes from the endpoint, $\tau=0$.

In the above procedure, the contour integration picks up the residues of the integrands at the simple poles $k=k_d$ (diffusive mode) and $k=k_{pl}$ (plasmon) in the upper half $k$-plane. The respective residue contributions for $E_r$ and $E_\phi$ on the sheet ($z=0$) in cylindrical coordinates are:
\begin{gather*}
E_r^d(r,0)=-\frac{\pi}{\omega\e}\,k_d^2 H_1^{(1)}(k_d r)\,\frac{\mathcal A(k_d)}{\mathcal D'(k_d)},\\
E_r^{pl}(r,0)=-\frac{\pi}{\omega\e}\, k_{pl}^2H_1^{(1)}(k_{pl} r)\frac{\mathcal A(k_{pl})}{\mathcal D'(k_{pl})},\\
E_{\phi}^d(r,0)=-\frac{\im\pi\omega^2 D_0}{c^2\e}\, k_d^2 H_1^{(1)}(k_d r)\frac{\omega_c+\eta_H k_d^2}{\mathcal D'(k_d)},\\
E_{\phi}^{pl}(r,0)=-\frac{\im\pi\omega^2 D_0}{c^2\e}\, k_{pl}^2 H_1^{(1)}(k_{pl} r)\frac{\omega_c+\eta_H k_{pl}^2}{\mathcal D'(k_{pl})}.
\end{gather*}
Here, the prime denotes differentiation with respect to the argument.

Hence, the tangential electric field components on the sheet ($z=0$) are given by
\begin{align*}
	E_{s}(r,0)=E^{bp}_s(r,0)+E^d_s(r,0)+E^{pl}_s(r,0);\qquad s=r,\,\phi.
\end{align*}
An interesting question is whether any particular residue contribution, i.e., the diffusive mode or the plasmon, can possibly be dominant in the above sum for some range of distance $r$. We investigate this issue numerically below.

\subsection{Numerical results }
\label{subsec:ampl_num}

Next, we numerically compute the radial and angular components of the electric field on the conducting sheet, using the techniques of Sec.~\ref{subsec:method}. We assess the relative importance and possible appearance of the collective modes, in comparison to the radiation field (i.e., the branch point contribution). In all computations, we assume that the ambient medium is the vacuum, thus setting $\e=1$.

\begin{figure}[h]
\begin{minipage}[h]{0.49\linewidth}
\center{\includegraphics[width=1\linewidth]{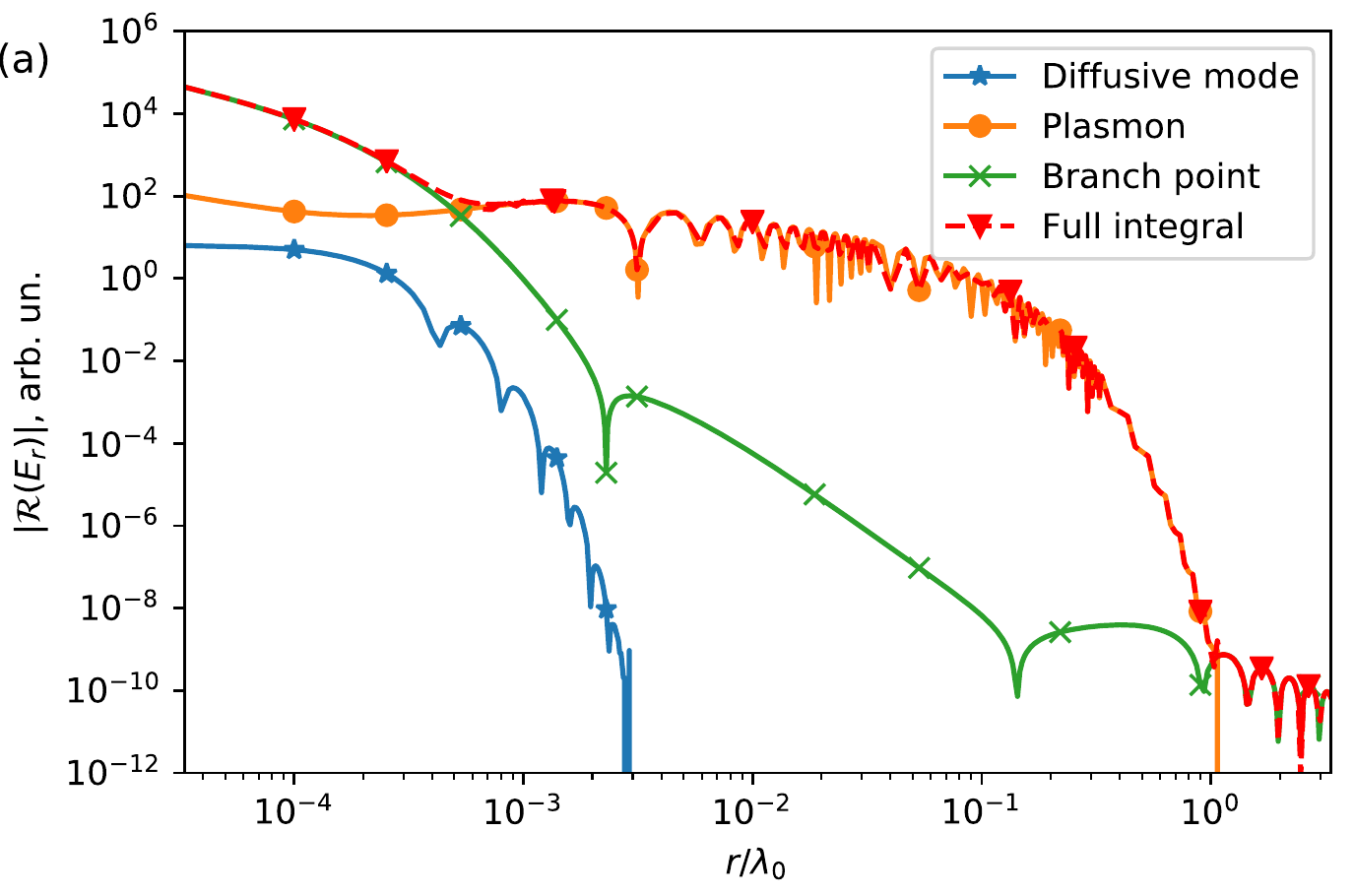}}
\end{minipage}
\hfill
\begin{minipage}[h]{0.49\linewidth}
\center{\includegraphics[width=1\linewidth]{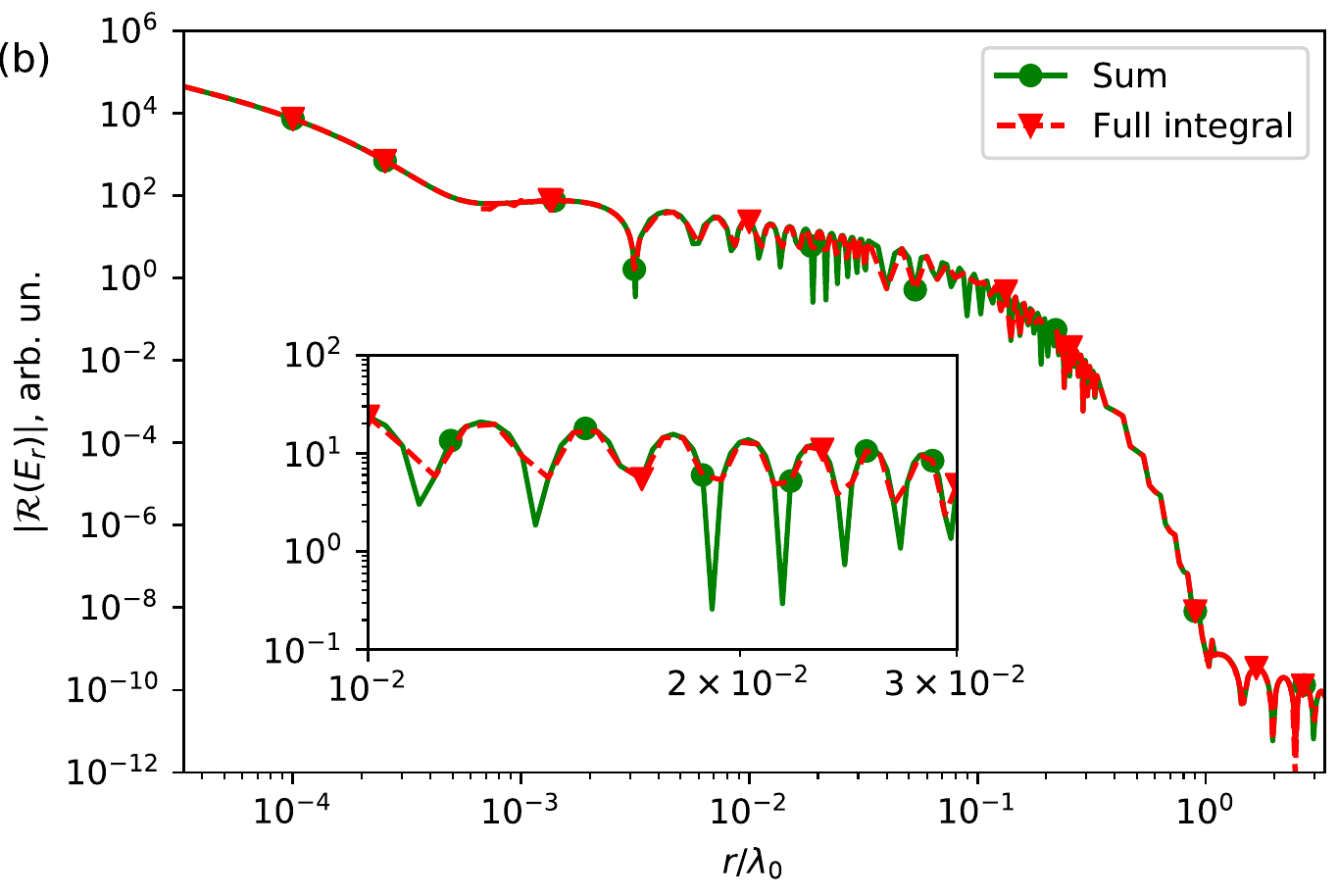}}
\end{minipage}
\vfill
\begin{minipage}[h]{0.49\linewidth}
\center{\includegraphics[width=1\linewidth]{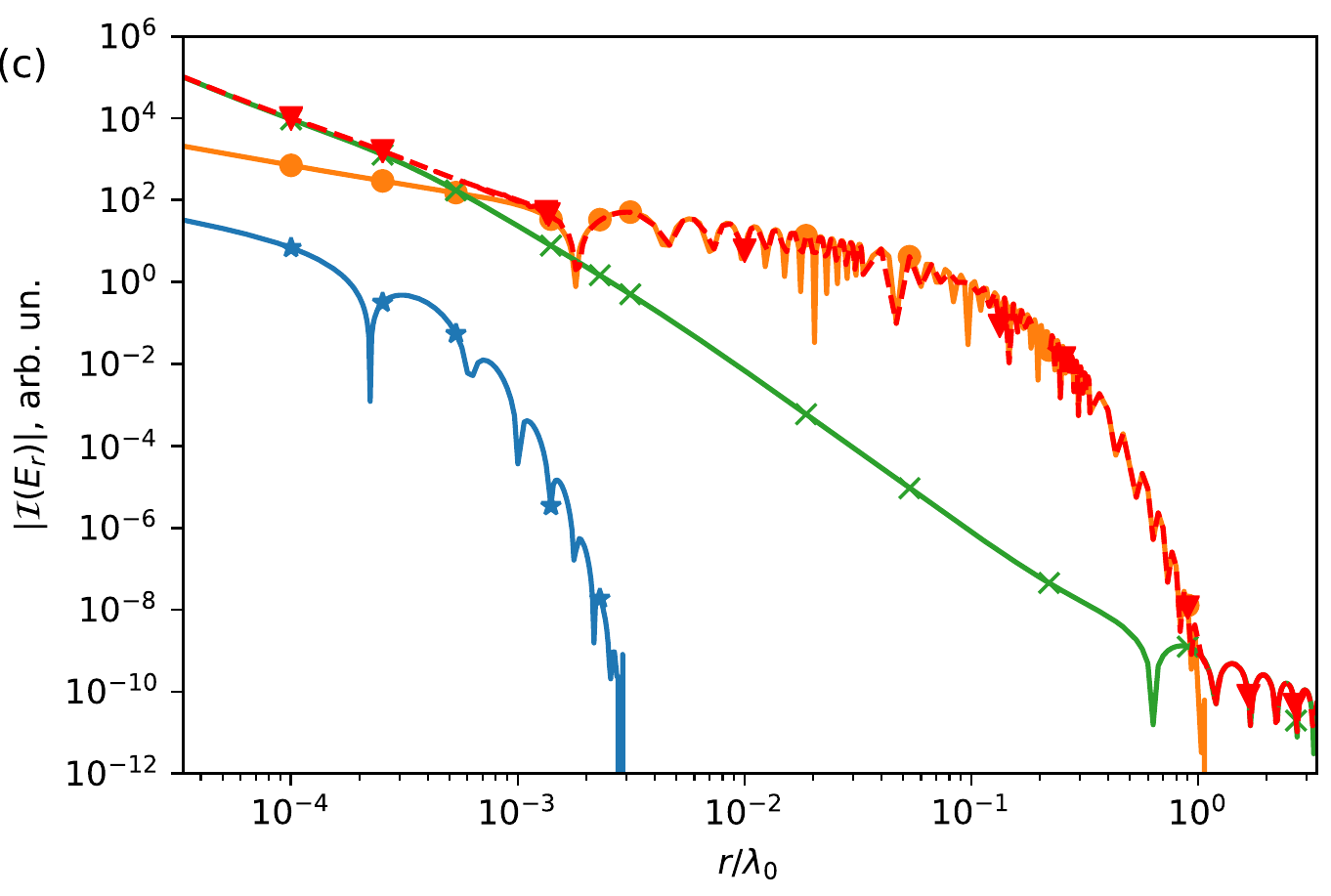}}
\end{minipage}
\hfill
\begin{minipage}[h]{0.49\linewidth}
\center{\includegraphics[width=1\linewidth]{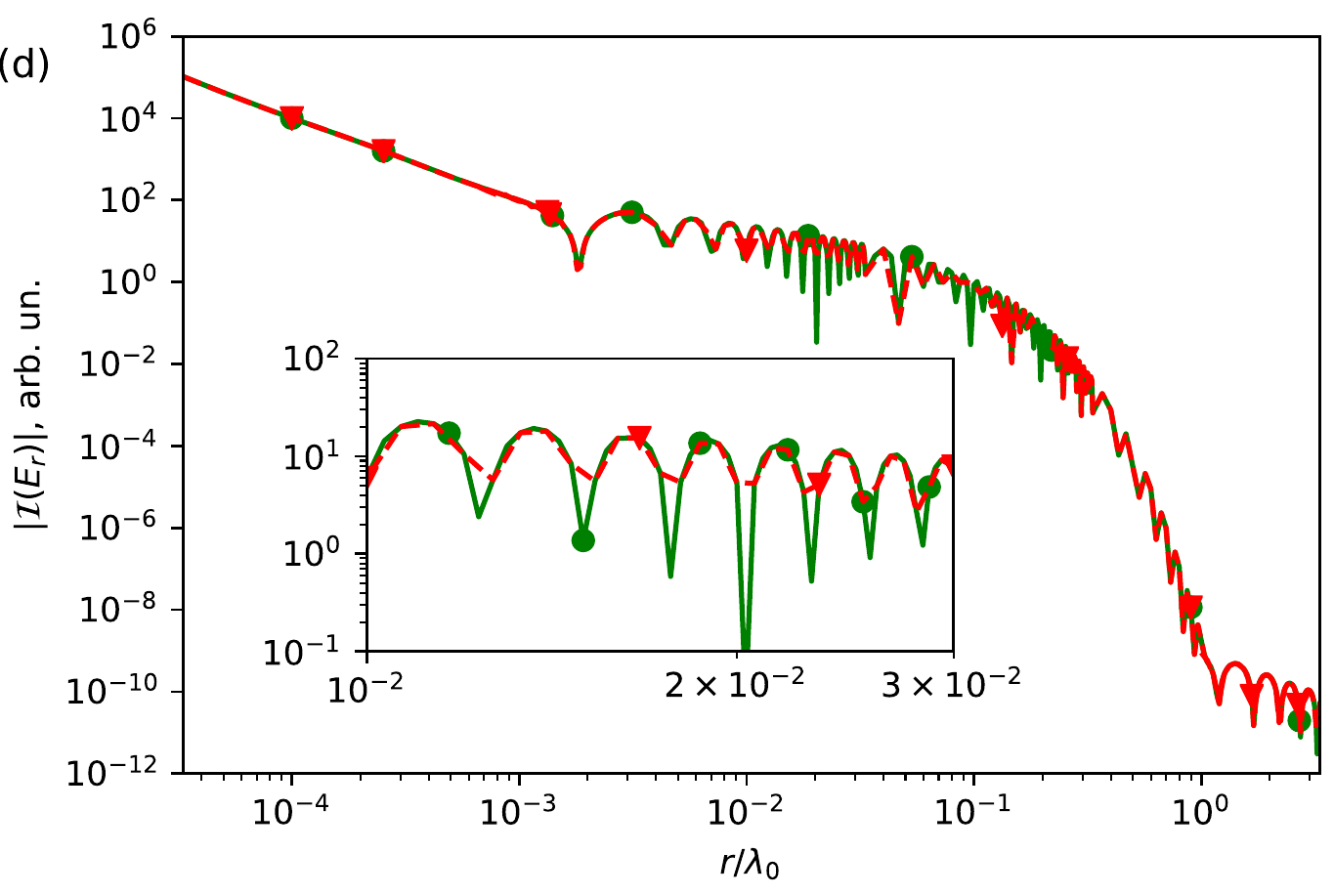}}
\end{minipage}
\centering{}\caption{(Color online) Absolute values of real [(a), (b)] and imaginary [(c), (d)] parts of the radial component $E_r(r,0)$ of the electric field generated by a Hertzian electric dipole on a graphene sheet. Left panel [(a), (c)]: Contributions from the diffusive mode (blue curve), plasmon (orange curve), and branch point (green curve), along with the full integral obtained by Eq.~\eqref{int_r} through direct numerical integration (red curve). Right panel [(b), (d)]: The same total electric field (red line) is compared to the one obtained as the sum of contributions from the residues of the two modes and the branch cut integral.
The numerical values of parameters used in the computations are: $\omega/(2\pi) = 1\,$ THz, $\gamma=0.01\,$THz, $n_0=10^{12}\,$cm$^{-2}$, $s=0.7\times10^5\,$m s$^{-1}$, $\eta_0(\omega/(2\pi)= 1\text{ THz}) \approx 0.004\,$ m$^2$s$^{-1}$, $B_{\text{st}} = 0.1\,$T.}
\label{fig_methods_contributions}
\end{figure}

In Fig.~\ref{fig_methods_contributions}, we show the log-log plots of the real and imaginary parts of the electric field radial component, $E_r(r,0)$, on the sheet versus the polar distance $r$ according to the full Fourier-Bessel integral~\eqref{int_r}. These values are compared to the three contributions mentioned above (two hydrodynamic-mode residues and branch cut integral). For example, in Fig.~\ref{fig_methods_contributions}(a) we plot each individual contribution to $\Re E_r(r,0)$; see the blue, orange, and green curves for the diffusive mode, plasmon, and branch cut integral, respectively. In Fig.~\ref{fig_methods_contributions}(b), we depict the sum of these three contributions. Note that the ``dips'' in the log-log plots correspond to harmonic oscillations with respect to $r$ that come from the first-kind Hankel function involved in the residues.

Our comparisons indicate that the \emph{near-field-region oscillations} ($k_0r\lesssim10^{-3}$) of the electric field should be attributed solely to the manifestation of the \emph{diffusive mode}. However, we observe that the amplitude of this mode dissipates quickly with $r$ on the sheet, as we analytically predict via its dispersion relation in Sec.~\ref{sec:poles}. In contrast, the \emph{plasmon oscillations} manifest in the \emph{intermediate-field region} ($10^{-3}\lesssim k_0r\lesssim1$) of the dipole source where the other two contributions are negligible. In fact, we notice that the plasmon mode dominates the total radial field component for a wide range of distances. However, if $k_0r$ is sufficiently large compared to unity ($k_0 r\gg 1$), both the plasmon and diffusive mode have substantially decayed. Consequently, by our numerics the \emph{two hydrodynamic modes are spatially separated}, because of their different wavelengths and dissipation rates. Specifically, by comparing the propagation length of the diffusive mode $l_{\text{d}}=1/k_d''$ and the plasmon wavelength $\lambda_{\text{pl}}=2\pi/k_{\text{pl}}'$ we find that the condition for the spatial separation, $l_{\text{d}}<\lambda_{\text{pl}}$, is satisfied for frequencies $\omega<\left(2\pi D_0/\sqrt{2\eta}\right)^{2/3}$. Accordingly, for the parameters considered in this paper, the frequency $\omega/(2\pi)$ should not exceed 4.6 THz. Hence, in principle, these modes might be observed (and distinguished) in a single experiment~\cite{Khavronin2020singularity}.

The numerical computations depicted in Figs.~\ref{fig_methods_contributions}(b), (d) aim to validate our two methods of integral evaluation (see Sec.~\ref{subsec:method}):  The red curve corresponds to the direct numerical evaluation of the integral given by
Eq.~\eqref{int_r}; and the blue curve amounts to the sum of the three individual contributions, which include the branch cut term. The two methods of evaluating $E_r$ are found to be in good agreement. This implies the mutual consistency, and plausible validity, of our numerical approaches.

\onecolumngrid

\twocolumngrid

\begin{figure*}[ht]
\includegraphics[width=1\linewidth]{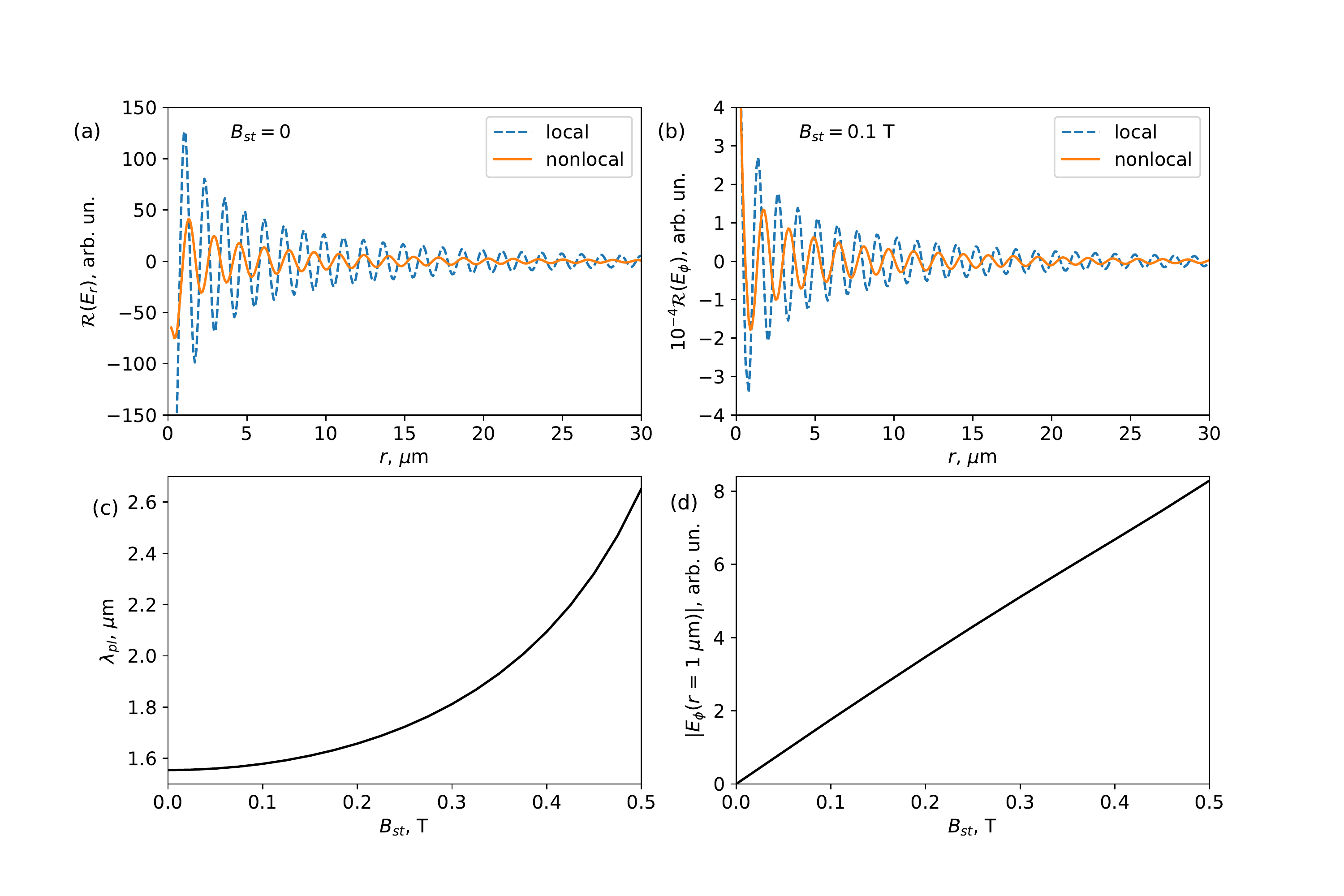}
\caption{(Color online) Real parts of radial component ($E_r$) and angular component ($E_\phi$) of the electric field on the sheet as functions of polar distance $r$ from the dipole source [(a), (b)]; plasmon wavelength versus static magnetic field [(c)]; and magnitude of angular component ($E_\phi$) of electric field at a fixed distance $r$ [(d)]. Top panel [(a), (b)]:
 Real part of component $E_r$ for zero static magnetic $B_{\text{st}} = 0\,$T  in the cases with local (blue curve) and nonlocal (orange curve) surface conductivity [left plot, (a)]; and corresponding real part of component $E_\phi$ for $B_{\text{st}}=0.1\,$T [right plot, (b)]. Bottom panel [(c), (d)]: Depiction of dependence of plasmon wavelength $\lambda_{pl}=2\pi/k'_{pl}$ on static magnetic field $B_{\text{st}}$ by dispersion relation~\eqref{eq:dr} [left plot, (c)]; and magnitude of component $E_\phi$ at polar distance $r=1\,\mu$m as a function of $B_{\text{st}}$ [right plot, (d)].
The numerical values of parameters used in the computations are: $\omega/(2\pi) = 1\,$ THz, $\gamma=0.01\,$THz, $n_0=10^{12}\,$cm$^{-2}$, $s=0.7\times10^5\,$m s$^{-1}$, $\eta_0(\omega/(2\pi)= 1\text{ THz}) \approx 0.004\,$ m$^2$s$^{-1}$.}
\label{fig_loc_vs_nonloc}
\end{figure*}

We now turn our attention to capturing nonlocal effects of the surface conductivity tensor (cf. Sec.~\ref{sec:conductivity}) on the excited electric fields. We remind the reader that, in the context of our hydrodynamic model, the local effects are brought about in the linear response of the 2D material regardless of the value of the static magnetic field when the compressional wave and all viscosities are switched off. This special case occurs when $s=0$ and $\eta_0=0$. Otherwise, the nonlocal effects of the compressional wave or viscosities are felt by the surface conductivity tensor. In Figs.~\ref{fig_loc_vs_nonloc}(a), (b) we indicate the influence of such nonlocal effects by plotting the real parts of the components $E_r$ [Fig.~\ref{fig_loc_vs_nonloc} (a)] and $E_\phi$ [Fig.~\ref{fig_loc_vs_nonloc} (b)] on the sheet as functions of the polar distance $r$ from the dipole source. Our numerical simulations show that both $E_r$ and $E_\phi$ dissipate more rapidly along the sheet when the nonlocal effects are present, in comparison to the setting of the Drude-like, local linear response. At the same time, the wavelength of the plasmon spatial oscillations increases. Note that this plasmon wavelength also increases with the static magnetic field $B_{\text{st}}$, as shown directly in Fig.~\ref{fig_loc_vs_nonloc}(c). In fact, this trend is in agreement with Fig. \ref{fig:disp_rel}(a). Furthermore, by Fig.~\ref{fig_loc_vs_nonloc}(d) we numerically assert that the magnitude of $E_\phi$ at a fixed distance $r$ of the order of the plasmon wavelength is roughly proportional to the static magnetic field $B_{\text{st}}$, for sufficiently weak $B_{\text{st}}$. This observation is compatible with integral representation~\eqref{int_phi} for $E_\phi$. Indeed, in this formula the numerator of the integrand is linear in the parameters $\omega_c$ and $\eta_H$ while the denominator, $\mathcal D(k)$, approaches a well-defined function of $k$ as $B_{\text{st}}$ becomes small enough. Recall that the Hall viscosity $\eta_H$ is approximately linear with $B_{\text{st}}$ if $|B_{\text{st}}|\ll B_0$ (see Sec.~\ref{sec:model}).

\onecolumngrid

\twocolumngrid

By dispersion relation~\eqref{eq:dr} and our numerics, the wavelengths of the plasmon and the diffusive mode can in principle be comparable at some frequency range. Therefore, it is challenging to try to distinguish the contributions of the two modes in a laboratory experiment. A plausible scenario for addressing this issue, as suggested by our analysis, is to excite the two modes by a vertical electric dipole operating at a relatively low frequency, say, $\nu=\omega/(2\pi)=0.1\,$THz, on the conducting sheet in the presence of a static magnetic field (perpendicular to the 2D material). In this setting, the static magnetic field can be adjusted to suppress the plasmon and thus single out the diffusive mode in the near field.

\begin{figure*}[ht]
\includegraphics[width=1\linewidth]{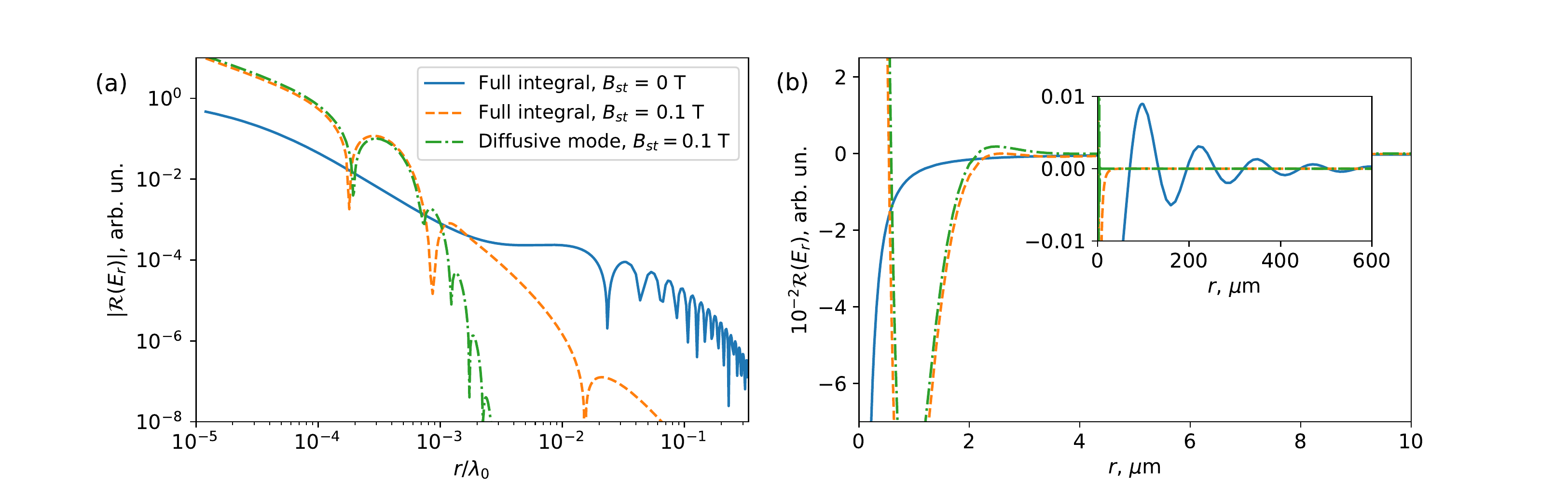}
\caption{(Color online) Numerical computation that indicates possible isolation of diffusive mode in radial component, $E_r$, of electric field under the influence of a static magnetic field ($B_{\text{st}}\bold{e}_z$) perpendicular to the sheet. Left [(a)]: Absolute value of the real part of $E_r$ as a function of radial distance $r$ from dipole source, in the near- and intermediate-field regions ($r$ is scaled by wavelength $\lambda_0$ in free space). Right [(b)]: Real part of $E_r$ as a function of radial distance $r$ in the near-field region. In both plots, the results from the full Fourier-Bessel integral (blue and orange curves) are plotted separately from the residue contribution of the diffusive mode (green curve). The static magnetic field values are $B_{\text{st}}=0\,$T (blue curve) and $B_{\text{st}}=0.1\,$T (orange and green curves). The plasmon manifests via fast spatial oscillations in the case with zero static magnetic field ($B_{\text{st}}=0$); in contrast, for $B_{\text{st}}=0.1\,$T, the plasmon becomes gapped and disappears. The numerical values of parameters used in the computations are: $\omega/(2\pi) = 0.1\,$ THz, $\gamma=0.01\,$THz, $n_0=10^{12}\,$cm$^{-2}$, $\eta_0(\omega/(2\pi)= 0.1\text{ THz}) \approx 0.2\,$ m$^2$s$^{-1}$ and $s=0.7\times10^5\,$m s$^{-1}$.}
\label{fig_B_vs_noB}
\end{figure*}

Figure~\ref{fig_B_vs_noB} illustrates the above scenario via the numerically evaluated real part of the electric-field radial component, $E_r$, as a function of the polar distance $r$ from the dipole source at frequency $\nu = 0.1\,$THz. We consider the cases without and with a static magnetic field ($B_{\text{st}}=0,\,0.1\,$T). In this plot, we display the respective values of the full Fourier-Bessel integral~\eqref{int_r} in comparison to the residue contribution of the diffusive mode. Note that the plasmon might manifest through the fast spatial oscillations of the full integral for the $\Re E_r$ in the intermediate-field region. We observe that the plasmon becomes gapped and disappears when a suitable value of the static magnetic field is applied, if $\omega_c > \omega=2\pi\nu$. Thus, the diffusive mode can, in principle, be isolated and detected in the near-field region if the value of the static magnetic field, $B_{\text{st}}$, and operating frequency, $\nu$, are such that the plasmon is gapped; see Fig.~\ref{fig_B_vs_noB}(b).

\begin{figure*}[ht]
\includegraphics[width=1\linewidth]{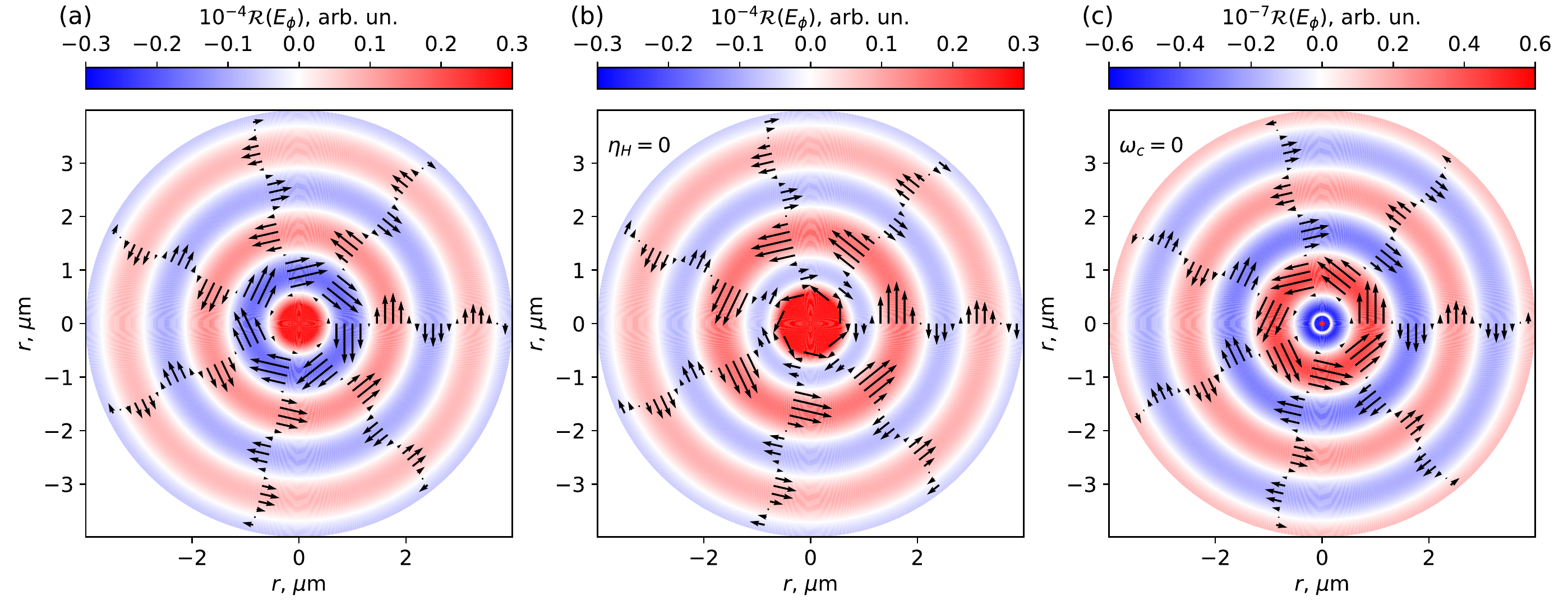}
\caption{(Color online) Two-dimensional representations of distinct contributions to the real part of the angular component, $E_{\phi}(r,0)$, of the electric field excited by a dipole source on a 2D material sheet, by numerical evaluation based on integral~\eqref{int_phi}. For comparison purposes, we separately turn off the effects of the Hall viscosity ($\eta_H$) and the Lorentz force ($\omega_c$) in our computations; cf. Eq.~\eqref{eq:hydro2}. (a) Real part of the total component $E_{\phi}(r,0)$. (b) Real part of $E_{\phi}(r,0)$ with zero Hall viscosity ($\eta_H=0$). (c) Real part of $E_{\phi}(r,0)$ with zero Lorentz force ($\omega_c=0$). The black arrows illustrate the vector field associated with $E_{\phi}$.
The numerical values of parameters used in the computations are: $\omega/(2\pi) = 1\,$ THz, $\gamma=0.01$ THz, $n_0=10^{12}$ cm$^{-2}$, $s=0.7\times10^5$ m s$^{-1}$, $\eta_0(\omega/(2\pi)= 1\text{ THz}) \approx 0.004\,$ m$^2$s$^{-1}$, $B = 0.01$ T.}
\label{fig_rings}
\end{figure*}

Before we close this section, it is of interest to discuss the contribution of the Hall viscosity to the angular component, $E_\phi$, of the electric field. This component deserves some special attention because, as we point out in Sec.~\ref{sec:integrals}, it vanishes identically only when $\omega_c=0$ and $\eta_H=0$ (thus, $B_{\text{st}}=0$); cf. Eq.~\eqref{eq:efi}. In Fig.~\ref{fig_rings}(a), we show the numerically evaluated $E_\phi$ by use of 2D color mappings. By comparing the wavelength of the oscillations and the distance, $r$, from the dipole source observed for $E_r$ in Fig. \ref{fig_rings}(a) to the corresponding quantities in Fig.~\ref{fig_methods_contributions}, we identify these oscillations with the plasmon. To study the effect of the Hall viscosity on the angular component, $E_\phi$, we turn off the Hall viscosity, setting $\eta_H=0$, and the Lorentz force, $\omega_c=0$, sequentially in our computations; cf. Eq.~\eqref{eq:hydro2}. The resulting color maps are shown in  Figs.~\ref{fig_rings}(b) and (c), respectively. Since we consider a weak magnetic field, $B_{\text{st}}=0.01$ T, the electric field shown in Fig. \ref{fig_rings}(a) is approximately equal to the sum of the electric fields depicted in Figs.\ref{fig_rings}(b), (c). In these plots, we can see that the contribution due to the Hall viscosity alone [Fig.~\ref{fig_rings}(c)] has an opposite sign compared to the Lorentz force contribution [Fig.~\ref{fig_rings}(b)]. This observation is consistent with previous studies performed by DC transport experiments~\cite{berdyugin2019measuring}.

Interestingly, the presence of the electric field angular component ($E_\phi$) spoils the longitudinal character of the plasmon oscillation. This means that the vector valued electric field, $\mathbf{E} = \mathbf{e}_r E_r+\mathbf{e}_{\phi} E_{\phi}+\mathbf{e}_z E_z$, now oscillates also in the direction perpendicular to the radial unit vector, $\mathbf{e}_r$, i.e., along the $z\phi$-cylindrical surface. To illustrate this feature, in all plots of Fig.~\ref{fig_rings} we indicate by black arrows the orientation of the constituent vector $\mathbf{e}_{\phi} E_{\phi}$.
At this point, it is worth pointing out that the diffusive mode remains transversal. This mode contributes to all three electric field components as dictated by the structure of our excitation dipole source in the presence of a static magnetic field.

\onecolumngrid

\twocolumngrid


\section{Discussion}
\label{sec:discussion}

In this section, we discuss and compare the features of hydrodynamic modes studied in Sec. V and VI. In particular, we outline possible implications of our results on the excitation of these modes by electric-current-carrying sources for future investigations in viscous 2DESs.

We should point out that both the diffusive mode and the plasmon are, in principle, sub-wavelength modes: their respective wavelengths can be much smaller than the wavelength of the ambient space at a suitable frequency regime. However, the diffusive mode is highly dissipative and can have a propagation length smaller than the wavelength of the plasmon. This property leads to the spatial separation of two hydrodynamic modes: the diffusive mode, having a slightly smaller wavelength, shows up closer to the excitation source and quickly decays, as it propagates farther away from the dipole, before the plasmon can manifest.

Although the two hydrodynamic modes can be spatially separated, as explained above, one can isolate the diffusive mode if needed. We described a scenario by which this isolation can be achieved by applying an external static magnetic field and choosing the frequency of the excitation source such that the plasmon is gapped. Unlike the plasmon, the diffusive mode (technically) exists for all frequencies even in the presence of the static magnetic field. In this case, the high-frequency oscillations in the resulting electric field in the 2DES can be attributed entirely to the diffusive mode.

Furthermore, we demonstrated that the momentum transfer between the layers of the electron liquid, which is usually carried out through the shear viscosity and is necessary for the existence of the diffusive mode, can in principle be done entirely by the Hall viscosity in the presence of the external static magnetic field. In this setting, one can possibly observe the Hall diffusive mode, which has a wavelength and propagation distance comparable to the ones of the regular diffusive mode. However, the dispersion relation of the Hall diffusive mode appears to be much more peculiar since it exhibits positive as well as \emph{negative} group velocities at different frequency ranges. This observation suggests that it will be worthwhile for a future effort to study the response of graphene to an excitation in the form of pulses (rather than monochromatic waves) in the viscous hydrodynamic regime of the 2DES.

Our analysis admits several tractable extensions and generalizations, which were not addressed in this paper. These extensions include the effect of the bulk viscosity in the numerical computations for the electric field, and the finite size of the 2DES when the electron fluid is confined in a channel. It will be worthwhile for a future effort to study the effect of the nonlinear material response on the diffusive mode and its extension beyond the frequency regime in which hydrodynamic theory applies~\cite{principi2019pseudo}.

\section{Conclusion}
\label{sec:conclusion}

In this paper, we described numerically and analytically the spatial structure of the electric field excited by a Hertzian electric dipole on an infinite, translation invariant sheet of a 2DES in the viscous hydrodynamic regime. To this end, we solved exactly a boundary value problem for the time-harmonic Maxwell equations coupled with linearized hydrodynamic (Navier-Stokes-type) equations for the 2D material. In our formalism, we took into account an external static magnetic field perpendicular to the sheet; and included the possible effects of the shear and Hall viscosities as well as the compressional wave of the 2DES. We placed particular emphasis on the amplitudes of collective modes that can be excited on the sheet in the far-infrared and THz frequency regimes. Our analysis singled out two types of modes, namely, the plasmon and diffusive mode. In the presence of an external static magnetic field perpendicular to the plane of the 2DES, the latter mode combines the features of both the conventional and Hall diffusion and may have a negative group velocity. We quantified the contributions of these modes relative to the radiation field by numerically evaluating Fourier-type integrals for the electric field tangential to the sheet.

By linear response theory, we also derived explicit formulas for the matrix elements of the resulting nonlocal surface conductivity tensor. In this description, the nonlocality comes from the effects of shear and Hall viscosities as well as that of the compressional wave. By calculating the Fourier-type integrals for the tangential electric field components, we indicated a scenario of separating the two collective modes at a suitable range of frequencies. We found that the plasmon may dominate in the intermediate-field region of the dipole source. In contrast, the diffusive mode prevails in the near-field region.

\acknowledgements
V.A., M.L. and D.M. acknowledge partial support by the ARO MURI Award W911NF-14-1-0247 and the Institute for Mathematics and its Applications (NSF Grant DMS-1440471) at the University of Minnesota for several visits. V.A's and M.L.'s research was also supported in part by NSF Awards DMS-1819220 and DMS-1906129. D.A.B. acknowledges support from the MIT Pappalardo Fellowship. The research of D.M. was also partially supported by a Research and Scholarship award by the Graduate School, University of Maryland, in the spring of 2019, when this work was initiated. Part of this research was carried out when three of the authors (V.A, M.L. and D.M.) were visiting the Institute for Pure and Applied Mathematics (IPAM), which is supported by NSF under grant DMS-1440415.

\begin{appendix}
\section{Integral representations for electric field}
\label{sec:derivation_E}

In this appendix, we derive the Fourier-Bessel integral representations
for the electric field components; cf. Eq.~\eqref{eqs:E-FB} in Sec.~\ref{sec:integrals}. The starting point is the boundary value problem for the time-harmonic Maxwell equations in the presence of the sheet by use of the surface conductivity tensor of Eq.~\eqref{eqs:sigma-tensor}.

Because of the sheet translation invariance, let
\begin{equation*}
\bold{H}\left(\br,z\right)=\frac{1}{4\pi^2}\iint d\bk\ \hat{\bold{H}}\left(\bk,z\right)\, e^{\im \bk\cdot \br},
\end{equation*}
where $\hat{\bold{H}}$ denotes the Fourier transform of the magnetic field; $\br=(x,y)$, $\bk=(k_x, k_y)$ and the integration range in the above Fourier integral is the entire $k_x k_y$-plane. Accordingly, we transform the curl laws of Maxwell's equations with respect to $x$ and $y$.
By solving the transformed Eq.~\eqref{eq:m1} for $\hat{\bold{E}}_j$ ($j=1,\,2$) and substituting the result into the transformed Eq.~\eqref{eq:m2}, we obtain the following differential equations (in the $z$ coordinate) for the tangential magnetic field components:
\begin{gather*}
(\partial_z^2-\beta^2)\hat{H}_{jx}=-\frac{4\pi \im}{c}k_y\,\delta(z-z_0),\\
(\partial_z^2-\beta^2)\hat{H}_{jy}=\frac{4\pi \im}{c}k_x\,\delta(z-z_0),
\end{gather*}
where $\beta^2=k^2-k_0^2$ and $k_0=\omega\sqrt{\e}/c$; $\partial_z=\partial/\partial z$. Here, we have set the dipole electric moment, $I_0\ell$, equal to unity. By symmetry, the magnetic field component perpendicular to the sheet vanishes identically, viz., $\hat{H}_{jz}\equiv 0$ for $j=1,2$ for all $z$. Note that the real part of $\beta$ is assumed to be positive ($\Re\beta(k)>0$).

The solutions of the above differential equations for $\hat{H}_{jx}(\bk, z)$ and $\hat{H}_{jy}(\bk, z)$ must decay with respect to $|z|$, away from the sheet ($z=0$). Therefore, we obtain the expressions
\begin{gather}
\label{hx}
\hat{H}_x=
\begin{cases}
\displaystyle K_{1x}e^{-\beta z}+\frac{2\pi \im}{c}\frac{k_y}{\beta}e^{-\beta\, \left|z-z_0\right|} & \text{for}\enspace z>0,
\\
K_{2x}e^{\beta z} & \text{for}\enspace z<0,
\end{cases}\\
\label{hy}
\hat{H}_y=
\begin{cases}
\displaystyle K_{1y}e^{-\beta z}-\frac{2\pi \im}{c}\frac{k_x}{\beta}e^{-\beta\, \left|z-z_0\right|} & \text{for}\enspace z>0,
\\
K_{2y}e^{\beta z} & \text{for}\enspace z<0.
\end{cases}
\end{gather}
The integration constants $K_{j\alpha}$ ($j=1,\,2$ and $\alpha=x,\,y$) should be determined via the transmission boundary conditions across the sheet (at $z=0$); see below.

By Eq.~\eqref{eq:m1}, the components of the transformed electric field are given by
\begin{gather*}
\hat{E}_x(\bk,z)=\frac{\im c}{\omega\e\beta^2}\left\{\left(k_0^2-k_x^2\right)(\partial_z\hat{H}_y)+k_xk_y\,(\partial_z\hat{H}_x)\right\},\\
\hat{E}_y(\bk,z)=-\frac{\im c}{\omega\e\beta^2}\left\{\left(k_0^2-k_y^2\right)(\partial_z\hat{H}_x)+k_xk_y\,(\partial_z\hat{H}_y)\right\},\\
\hat{E}_z(\bk,z)=\frac{c}{\omega\e}\left(k_y\hat{H}_x-k_x\hat{H}_y\right),\qquad z\neq z_0.
\end{gather*}

To express the components of $\hat{\bold{E}}_j$ in terms of the integration constants $K_{j\alpha}$ ($j=1,\,2$ and $\alpha=x,\,y$), we now invoke Eqs.~\eqref{hx} and~\eqref{hy}. Hence, we rewrite the transformed electric field components as
\begin{gather*}
\label{ex}
\hat{E}_x=\frac{\im\omega}{ck_0^2\beta(k)}
\begin{cases}
\begin{split}
-\left\{k_xk_yK_{1x}+\left(k_0^2-k_x^2\right)K_{1y}\right\}e^{-\beta(k) z}\\
-\frac{2\pi \im}{c}k_x\beta(k) \sgn(z-z_0)\,e^{-\beta(k)\, \left|z-z_0\right|}
\enspace\text{for}\enspace z>0,
\end{split}
\\
\left\{k_xk_yK_{2x}+\left(k_0^2-k_x^2\right)K_{2y}\right\}e^{\beta(k)\, z}  \enspace\text{for}\enspace z<0,
\end{cases}\\
\label{ey}
\hat{E}_y=-\frac{\im\omega}{ck_0^2\beta(k)}
\begin{cases}
\begin{split}
-\left\{k_xk_yK_{1y}+\left(k_0^2-k_y^2\right)K_{1x}\right\}e^{-\beta(k)\, z}\\
+\frac{2\pi \im}{c}k_y\beta(k)\, \sgn(z-z_0)\,e^{-\beta(k)\, \left|z-z_0\right|} \enspace\text{for}\enspace z>0,
\end{split}
\\
\left\{k_xk_yK_{2y}+\left(k_0^2-k_y^2\right)K_{2x}\right\}e^{\beta(k)\, z} \enspace\text{for}\enspace z<0,
\end{cases}\\
\hat{E}_z=\frac{c}{\omega\e}
\begin{cases}
\begin{split}
\left(k_yK_{1x}-k_xK_{1y}\right)e^{-\beta(k)\, z}
+\frac{2\pi \im}{c}\frac{k^2}{\beta(k)} e^{-\beta(k)\, \left|z-z_0\right|}\\ \enspace\text{for}\enspace z>0, z\neq z_0,
\end{split}
\\
\left(k_yK_{2x}-k_xK_{2y}\right)e^{\beta(k)\, z} \enspace\text{for}\enspace z<0.
\end{cases}
\end{gather*}

The remaining task is to find the ($\bk$-dependent) coefficients $K_{j\alpha}$
($j=1,\,2$ and $\alpha=x,\,y$). To this end, we use boundary conditions~\eqref{eq:bc1} and~\eqref{eq:bc2}, along with Eqs.~\eqref{eq:sigma-xx}-\eqref{eq:sigma-yy} for the matrix elements of the sheet tensor conductivity. After some algebra, we obtain the formulas
\begin{gather*}
K_{1x}=-\frac{2\pi \im D_0}{c\e}e^{-\beta z_0}\frac{\left(k_0^2 D_0+\beta\omega\Omega\right)k_y+i\beta\omega \Omega_c(k) k_x}{\mathcal D},\\
K_{2x}=\frac{2\pi \im}{c\beta}e^{-\beta z_0}\frac{-\mathcal A\,k_y+\im\beta^2 D_0\omega \Omega_c(k) k_x}{\mathcal D},\\
K_{1y}=\frac{2\pi \im D_0}{c\e}e^{-\beta z_0}\frac{\left(k_0^2 D_0+\beta\omega\Omega\right)k_x-\im\beta\omega\Omega_c(k) k_y}{\mathcal D},\\
K_{2y}=\frac{2\pi \im}{c\beta}e^{-\beta z_0}\frac{\mathcal A k_x+\im\beta^2 D_0\omega \Omega_c(k) k_y}{\mathcal D},
\end{gather*}
where

\begin{gather*}
\begin{split}
\mathcal A(k)&=\left[\left(s^2-\im\omega\zeta\right) k^2-\omega\Omega(k)\right]\left(k^2D_0+\e\omega\Omega(k)\beta(k)\right)\\
&+\omega^2\e\beta(k) \Omega_c(k)^2,
\end{split}\\
\begin{split}
\mathcal D(k)=&-\mathcal A(k)
-\beta(k) k^2 D_0^2/\e-D_0\omega\Omega(k)\beta^2(k),
\end{split}\\
\Omega(k)=\omega+\im\gamma+\im\eta k^2,\\
\Omega_c(k)=\omega+\im\eta_H k^2.
\end{gather*}
In the above, $D_0=2\pi e^2 n_0/m$ is the Drude weight. The substitution of these expressions for $K_{j\alpha}$ into the formulas for the transformed electric field components and Fourier inversion yield double Fourier integrals for $E_{j\alpha}(x,y,z)$. The use of cylindrical coordinates $(r,\phi,z)$ instead of $(x,y,z)$ via the corresponding change of variables $k_x=k\cos\phi'$ and $k_y=k \sin \phi'$ ($k=(k_x^2+k_y^2)^{1/2}\ge 0$) in the double Fourier integrals then results in the Fourier-Bessel integral representations diplayed in Eq.~\eqref{eqs:E-FB}~\cite{LateralWaves-book}.

\end{appendix}

\bibliography{biblio-last}

\end{document}